\documentclass{emulateapj}
\usepackage{lscape}

\newcommand{\Ha}{\mbox{H$\alpha$}}
\newcommand{\Hb}{\mbox{H$\beta$}}
\newcommand{\Hd}{\mbox{H$\delta$}}
\newcommand{\Hg}{\mbox{H$\gamma$}}
   
\newcommand{\etal}{\mbox{et al.}}

\newcommand{\ergscms}{ergs cm$^{-2}$ s$^{-1}$}

\newcommand{\Te}{\mbox{$T_{\rm{e}}$}}
\newcommand{\Ne}{\mbox{$N_{\rm{e}}$}}

\newcommand{\Apone}{\mbox{Ap1}}
\newcommand{\Aptwo}{\mbox{Ap2}}
\newcommand{\Apthree}{\mbox{Ap3}}
\newcommand{\Apfour}{\mbox{Ap4}}
\newcommand{\Apfive}{\mbox{Ap5}}
\newcommand{\Apsix}{\mbox{Ap6}}

\newcommand{\Sone}{\mbox{S1}}
\newcommand{\Stwo}{\mbox{S2}}

\newcommand{\knotB}{\mbox{\textsc{b}}}
\newcommand{\knotBone}{\mbox{\textsc{b1}}}
\newcommand{\knotBtwo}{\mbox{\textsc{b2}}}
\newcommand{\knotA}{\mbox{\textsc{a}}}
\newcommand{\knotC}{\mbox{\textsc{c}}}
\newcommand{\knotD}{\mbox{\textsc{d}}}
\newcommand{\knotE}{\mbox{\textsc{e}}}

\newcommand{\Sersic}{\mbox{S{\'e}rsic}}

\newcommand{\BV}{\mbox{$B-V$}}
\newcommand{\VK}{\mbox{$V-K$}}
\newcommand{\VR}{\mbox{$V-R$}}
\newcommand{\BR}{\mbox{$B-R$}}
\newcommand{\VI}{\mbox{$V-I$}}
\newcommand{\VJ}{\mbox{$V-J$}}
\newcommand{\HK}{\mbox{$H-K$}}
\newcommand{\JH}{\mbox{$J-H$}}

\newcommand{\BVRI}{\hbox{$BV\!RI$}} 
\newcommand{\JHK}{\hbox{$J\!H\!K$}}

\begin{document}

\slugcomment{Accepted for publication in {\it The Astrophysical
Journal}}

\title{Spectrophotometric investigations of Blue Compact Dwarf Galaxies:  
Markarian~35}

\author{Luz M. Cair{\'o}s} 
\email{luzma@aip.de} 
\affil{Astrophysikalisches Institut Potsdam, An der Sternwarte 16, 
D-14482 Potsdam, Germany} 

\author{Nicola Caon}
\email{ncaon@ll.iac.es}
\affil{Instituto de Astrof{\'\i}sica de Canarias, E-38200  La Laguna, Tenerife,
Canary Islands, Spain}

\author{Bego\~na Garc{\'\i}a Lorenzo}
\email{bgarcia@iac.es}
\affil{Instituto de Astrof{\'\i}sica de Canarias, E-38200  La Laguna, Tenerife,
Canary Islands, Spain}

\author{Ana Monreal-Ibero}
\email{amonreal@aip.de}
\affil{Astrophysikalisches Institut Potsdam, An der Sternwarte 16, 
D-14482 Potsdam, Germany}

\author{Ricardo Amor{\'\i}n}
\email{ramorin@ll.iac.es}
\affil{Instituto de Astrof{\'\i}sica de Canarias, E-38200  La Laguna, Tenerife,
Canary Islands, Spain}

\author{Peter Weilbacher}
\email{pweilbacher@aip.de}
\affil{Astrophysikalisches Institut Potsdam, An der Sternwarte 16, 
D-14482 Potsdam, Germany}

\author{Polychronis Papaderos}
\email{p.papaderos@observational-cosmology.eu}
\affil{Instituto de Astrof{\'\i}sica de Andaluc{\'\i}a, Apdo. 3004, 
18080 Granada, Spain}

%%\accepted{}
\shortauthors{Cair{\'os} et al.}
\shorttitle{Spectrophotometric observations of Mrk~35}

\begin{abstract} 

We present results from a detailed spectrophotometric analysis of the blue
compact dwarf galaxy Mrk~35 (Haro~3), based on deep optical (\BVRI) 
and near-IR (\JHK) imaging, \Ha\
narrow-band observations and long-slit spectroscopy. The optical emission of 
the galaxy is dominated by a central young starburst, with a bar-like shape, 
while an underlying component of stars, with elliptical isophotes and red 
colors, extends more than 4 kpc from the galaxy center. High resolution \Ha\ 
and color maps allow us to identify the star-forming regions, to spatially 
discriminate them from the older stars, and to recognize several dust patches. 
We derive colors and \Ha\ parameters for all the identified star-forming knots.
Observables derived for each knot are corrected for the contribution of the
underlying older stellar population, the contribution by emission lines, and
from interstellar extinction, and compared with evolutionary synthesis models.
We find that the contributions of these three factors are by no means 
negligible and that they significantly vary across the galaxy.
Therefore, careful quantification and subtraction of emission lines, galaxy
host contribution, and interstellar reddening at every galaxy position, are
essential to derive the properties of the young stars in BCDs. We find that we 
can reproduce the colors of all the knots with an instantaneous burst of star
formation and the Salpeter initial mass function with an upper mass limit of 
100 $M_{\sun}$. In all cases the knots are just a few Myr old. The underlying
population of stars has colors consistent with being several Gyr old.

\end{abstract}


\section{INTRODUCTION}

Blue Compact Dwarf (BCD) galaxies are narrow emission-line dwarfs that
are undergoing violent bursts of star-formation (Sargent
\& Searle 1970). They are compact and low-luminosity objects (starburst
diameter $\leq 1$ kpc; $M_{B} \geq -18$ mag), often with low-metal content 
($Z_{\odot}/50 \leq Z \leq Z_{\odot}/2$) and high star-forming (SF) rates, 
able to exhaust their gas on a time scale much shorter than the age 
of the Universe. Initially it was hypothesized that BCDs were truly young 
galaxies, forming their first generation of stars \citep{Sargent70, 
Lequeux80, Kunth86, Kunth88}, but the subsequent detection of an extended, 
redder stellar host in the vast majority of them has shown that most 
BCDs are old systems (\citealt{LooseThuan86, Telles95, Papaderos96a, 
Cairos00}; 
\citealt[][=C01a]{Cairos01a}; 
\citealt[][=C01b]{Cairos01b}; 
\citealt[][=C02]{Cairos02};
\citealt[][=C03]{Cairos03}; \citealt{BergvallOstlin02}). 
Although so far the SF in BCDs has been commonly described as a bursting 
scenario---short intense episodes of SF separated by long inactivity periods 
\citep{Thuan91, Mashesse99}---, there is increasing evidence that 
BCDs could rather have a gasping star formation, with long episodes of 
activity, separated by short quiescent intervals if any \citep{Legrand2000, 
Legrandetal2000, RecchiHensler2005}.

The evolutionary status of BCDs and their star formation history, as well as
the  mechanisms that trigger the recurrent star formation episodes, are still
open questions, and, although much work has been done in the last years in
this topic, no conclusive results have been achieved.

A fundamental step to approach such questions is a detailed  analysis of
individual nearby objects, using high quality data. In fact, most of the work
done so far has focused on statistical analyses of BCD samples, and only
recently a few studies have been devoted to examining in detail the
characteristics of individual objects. \cite{Papaderos98,Papaderos99} carried
out a spectrophotometric analysis of  SBS~0335-052 and Tololo~65, two BCDs
belonging to the i0 class (extremely  compact and low-metallicity objects with
no evidence for a substantial underlying stellar component; see
\citealt{LooseThuan86} for the classification scheme); \cite{Noeske00} studied
the i,IC galaxies Mrk~59 and Mrk~71 ("Cometary BCDs", galaxies with star
formation concentrated on one side);  Tololo~1214-277 and SBS 0940+544 (also
classified as i,IC) were studied  by \cite{Fricke01} and \cite{Guseva01},
respectively. Despite the fact that nE (objects with a single central
starburst superposed on a regular redder outer envelope) and iE (objects with 
a complex  inner  structure consisting of many SF knots atop a regular redder
outer envelope)  BCDs are the most common, there are very few studies of 
similar quality for  galaxies belonging to these groups, the most notable one 
being the comprehensive analysis of the properties of the iE Mrk~86 carried 
out by \cite{GildePaz00a} and \cite{GildePaz00b}.

This prompted us to start an extensive observational project, whose aim is the
thorough analysis of a sample of nearby BCD galaxies selected among those
already studied by our group (C01a and C01b). In a first paper (C02) we
introduced the method used to discriminate and analyze the stellar populations
in BCDs, and  presented the results for the iE galaxy Mrk~370.  Here we apply
the same method to Mrk~35, a BCD also belonging to the iE morphological class.

Mrk~35 has an absolute magnitude $M_{B} = -17.75$ (C01b), and is located at a
distance of 15.6 Mpc\footnote{This distance has been computed assuming a Hubble
flow, with a Hubble constant $H_0 = 75$ km  sec$^{-1}$ Mpc$^{-1}$ and applying
the correction for the Local group infall into Virgo.}.  According to the
morphology and distribution of its SF knots, Mrk~35 belongs  to the {\sc Type
iii} (Chain/Aligned starburst) group, as introduced in C01b.  The basic data of
Mrk~35 are shown in Table~\ref{Table:data}\footnote{A  collection of color maps
of this galaxy can be found at: \url{http://www.iac.es/proyect/GEFE/BCDs/}}. 

Deep optical surface photometry, in the optical and in the NIR, was presented 
in C01ab and in C03. The galaxy displays the typical surface brightness profile
(SBP) of BCD galaxies: at high and intermediate intensity levels the profile is
dominated by the starburst component, which has quite a complex shape, whereas
in the outer parts the SBP traces the luminosity structure of the old stellar
population. Mrk~35 is included in the \cite{MazBor93} sample of Markarian
galaxies with multiple nuclei. The presence of Wolf-Rayet stars
\citep{Steel96,Huang99} indicates that the galaxy underwent a starburst episode
within the last 3-6 Myr. The cause of this star-formation episode is unclear.
Whereas \cite{Steel96} conclude that the star-formation is likely auto-induced,
\cite{Sanchez00} claim that Mrk~35 shows clear signatures of a merger episode. 
A recent paper by \cite{Johnson04} studies the central starburst by means of 
near infrared and radio observations, and concludes that the hypothesis of a 
small scale interaction can not be ruled out.

\begin{deluxetable}{lccccccc}
\tabletypesize{\footnotesize}
\tablewidth{0pt}
\tablecaption{Collection of data on Mrk~35}
\tablehead{
  \colhead{Parameter}  & \colhead {Data} & \colhead {Reference} }
\startdata
 Other names      & Haro~3, NGC~3353         &     \\
 R.A. (J2000)     &   10 45 22               &     \\
 Decl. (J2000)    &   55 57 37               &     \\
 $D$ (Mpc)        &  15.6                    & (1) \\
 $A_{B}$          &  0.031                   & (2) \\
 $M_{B}$          & $-17.75$                 & (3) \\
 $m_{B}$          &  \phs13.21               & (3) \\
 $m_{V}$          &  \phs12.60               & (3) \\
 $m_{R}$          &  \phs12.37               & (3) \\
 $m_{I}$          &  \phs11.84               & (3) \\
 $m_{J}$          &  \phs11.31               & (4) \\
 $m_{H}$          &  \phs10.83               & (4) \\
 $m_{K_{s}}$      &  \phs10.43               & (4) \\
 $M_{HI}$         &  0.46$\times$10$^{9}$    & (5) \\
 $M_{T}$          &  $1.5\times 10^{9}$      & (5) \\
\enddata
\label{Table:data}
\tablecomments{(1) Distance computed assuming a Hubble flow, with a Hubble 
constant $H_0 = 75$ km sec$^{-1}$ Mpc$^{-1}$, and applying the correction 
for the Local Group infall into Virgo.
(2) Absorption coefficient in the $B$ band, from \cite{Schlegel98}.
(3) Asymptotic photometry obtained by extrapolating the growth curves, and
corrected for Galactic extinction \citep{Cairos01b}.
(4) Isophotal magnitudes (within the isophote at 23.0 mag arcsec$^{-2}$ in 
$J$, 22.0 in $H$ and 21.0 in $K_s$; C03).
(5) Neutral hydrogen mass $M_{HI}$ and total mass $M_{T}$ in units of 
$M_\odot$; both from \citet{ThuanMartin81}. Units of right ascension are
hours, minutes, and seconds, and units of declination are degrees, arcminutes,
and arcseconds.}
\end{deluxetable}

\section{OBSERVATIONS AND DATA REDUCTION}

\subsection{Long-slit spectra} 

Long-slit spectra of Mrk~35 were obtained in February 2002 at the
Observatorio del Roque de Los Muchachos (ORM) on the island of La Palma, using
the 4.2m William Herschel Telescope (WHT). Observations were made
with the blue arm of the ISIS double beam spectrograph, which was equipped 
with a 300 groove mm$^{-1}$ grating and a CCD array of $2100\times4200$  
$13.5\mu$ pixels, a combination which gives a linear dispersion of 0.86 \AA\ 
per pixel, and a spectral range of $3600$ to $6920$ \AA. 
A slit 4 arcmin long and $1\farcs2$ wide was used. The positions of the slit 
we used is shown in Fig.~\ref{Fig:slitpos}, over-plotted on a 
continuum-subtracted \Ha\ map.

The exact positions of the slit were derived from a direct image obtained from
the ISIS acquisition TV. The position angle of the two slit positions was set 
to $42\arcdeg$ and $38\arcdeg$ (S1 and S2 in Fig.~\ref{Fig:slitpos} 
and hereafter), so as to pass through the brightest central knots and the 
knots in the east tail. We took three exposures of 1800 seconds in the S1 
position, and two exposures of 1800 seconds in the S2 position. The average 
seeing during the observations was about $1\farcs2$. 

Data reduction was performed using IRAF standard tasks: after subtracting the 
bias and flat-fielding the spectra, they were calibrated in wavelength. 
The sky spectrum was derived by averaging the signal in two windows 60--70 
pixel wide outside the region where the object is still detectable, and 
subtracted out. 
The spectra were then corrected for atmospheric extinction and flux 
calibrated by means of observations of spectrophotometric standards.

\begin{deluxetable*}{ccccccc}
\tablecaption{Log of the observations \label{Tab:Obslog}}
\tablehead{
\colhead{Date} & \colhead{Telescope} & 
            \colhead{Instrument} & \colhead{Filter/grism} 
          & \colhead{Exposure time (s)} &\colhead{Seeing}  }
\startdata
 Jan. 99 & NOT 2.56m & ALFOSC  & $B$        &     1800   & $1\farcs55$ \\
 Jan. 99 & NOT 2.56m & ALFOSC  & $V$        &     1200   & $1\farcs12$ \\
 Jan. 99 & NOT 2.56m & ALFOSC  & $R$        &  \phn960   & $1\farcs44$ \\
 Jan. 99 & NOT 2.56m & ALFOSC  & $I$        &     1200   & $1\farcs00$ \\
 Mar. 02 & NOT 2.56m & ALFOSC  & 6562 (46)  &     3600   & $1\farcs20$ \\
 Mar. 02 & NOT 2.56m & ALFOSC  & 6353 (45)  &     3600   & $1\farcs18$ \\
 Apr. 00 & WHT 4.2m  & INGRID  & $J$        &  \phn800   & $1\farcs00$ \\
 Apr. 00 & WHT 4.2m  & INGRID  & $H$        &  \phn720   & $1\farcs02$ \\
 Apr. 00 & WHT 4.2m  & INGRID  & $K_{s}$    &     1440   & $0\farcs90$ \\
 Feb. 02 & WHT 4.2m  &  ISIS   & R300B      & \phm{00000}5400 (S1) & $1\farcs20$ \\
 Feb. 02 & WHT 4.2m  &  ISIS   & R300B      & \phm{00000}3600 (S2) & $1\farcs20$ \\[4pt]
\enddata
\label{Table:log}
\tablecomments{NOT = Nordic Optical Telescope;
ALFOSC = Andalucia Faint Object Spectrograph and Camera;
WHT = William Herschel Telescope; 
INGRID = Isaac Newton Group Red Imaging Device;
ISIS = Intermediate dispersion Spectrograph and Imaging System.}
\end{deluxetable*}

\subsection{H$\alpha$ imaging}

Images centered on the \Ha\ line and adjacent continuum were acquired in March
2002 at the Nordic Optical Telescope (NOT; ORM), using  ALFOSC ({\em Andalucia
Faint Object Spectrograph and Camera}). The detector  was a 2k $\times$ 2k,
15\micron\ pixel Ford-Loral CCD, which, with a scale of 0.188 arcsec per pixel,
provides a field of view of $6\farcm4 \times 6\farcm4$. The seeing varied 
around $1\farcs2$. The complete log of the observations is shown in 
Table~\ref{Table:log}. 

The image reduction was carried out using IRAF. Each image was corrected for 
bias, using an average bias frame, and was flattened by division by a
mean twilight flatfield image. The average sky level was estimated by 
computing the mean value within several boxes surrounding the object, and 
subtracted out as a constant. The frames were then registered (for each 
filter we took a set of three dithered exposures) and combined to obtain 
the final frame, with cosmic ray events removed and bad pixels cleaned out. 

The \Ha\ frame contains the emission from both the \Ha\ emission line and 
the underlying continuum; in order to obtain the net \Ha\ we need to subtract 
the continuum flux (the off-line image) from the \Ha\ total emission. 
First, the sky-subtracted off-line frame was rescaled in intensity to match 
the continuum in the sky-subtracted \Ha\ frame. 
The scale factor was calculated by matching the intensities in the two 
frames of several bright, unsaturated field stars, which are assumed not to 
have \Ha\ emission. 
The accuracy of the scale factor, indicated by the scatter shown among the 
values for different stars, is $\leq$ 5\%. The reliability of the method 
is corroborated by the lack of systematic residuals in the stellar images in 
the net \Ha\ frame.

The next step was the calibration in flux of the continuum-subtracted (net) 
\Ha\ frame. This was done by using the observations of spectrophotometric 
stars selected from the lists of \citet{Oke90}. 
We observed stars {\sc Feige~34} and {\sc bd+33~2642}, whose flux is 
calibrated and tabulated in steps of $\leq$ 2\AA. This high resolution allows 
us to calculate their flux (or magnitude) through each one of the
narrow-band filters we used, by integrating through their transmission curves.

\subsection{Broadband Imaging}

Broad-band observations of Mrk~35 were carried out in January 1999, at the 
NOT, equipped with ALFOSC. We  collected CCD images through the $B$, $V$, $R$ 
and $I$ filters. The seeing varied  between $1\farcs0$ and $1\farcs5$. The log
of  the observations is shown in Table~\ref{Table:log}. The data reduction and
calibration is presented in C01a,b.

\subsection{Near Infrared Imaging}

NIR imaging of the sample galaxies was obtained in April 2000, at the 4.2-m 
WHT (ORM, La Palma). We used the infrared camera INGRID (Isaac Newton Group 
Red Imaging Device), equipped with a $1024\times1024$ Hawaii near-IR detector
array. The pixel scale is $0\farcs242$ pixel$^{-1}$, and the field of view
$4\farcm1 \times 4\farcm1$. We took images in the $J$, $H$ and  $K_{s}$ 
bands.  
The seeing was around 1 arcsec. A complete log of the observations is provided 
in Table~\ref{Tab:Obslog}.
The data reduction is described in C03.

\section{RESULTS} 

\subsection{Spectroscopic Analysis}

\begin{figure*}[h]   
\centerline{\hbox{
\includegraphics[width=3.40in]{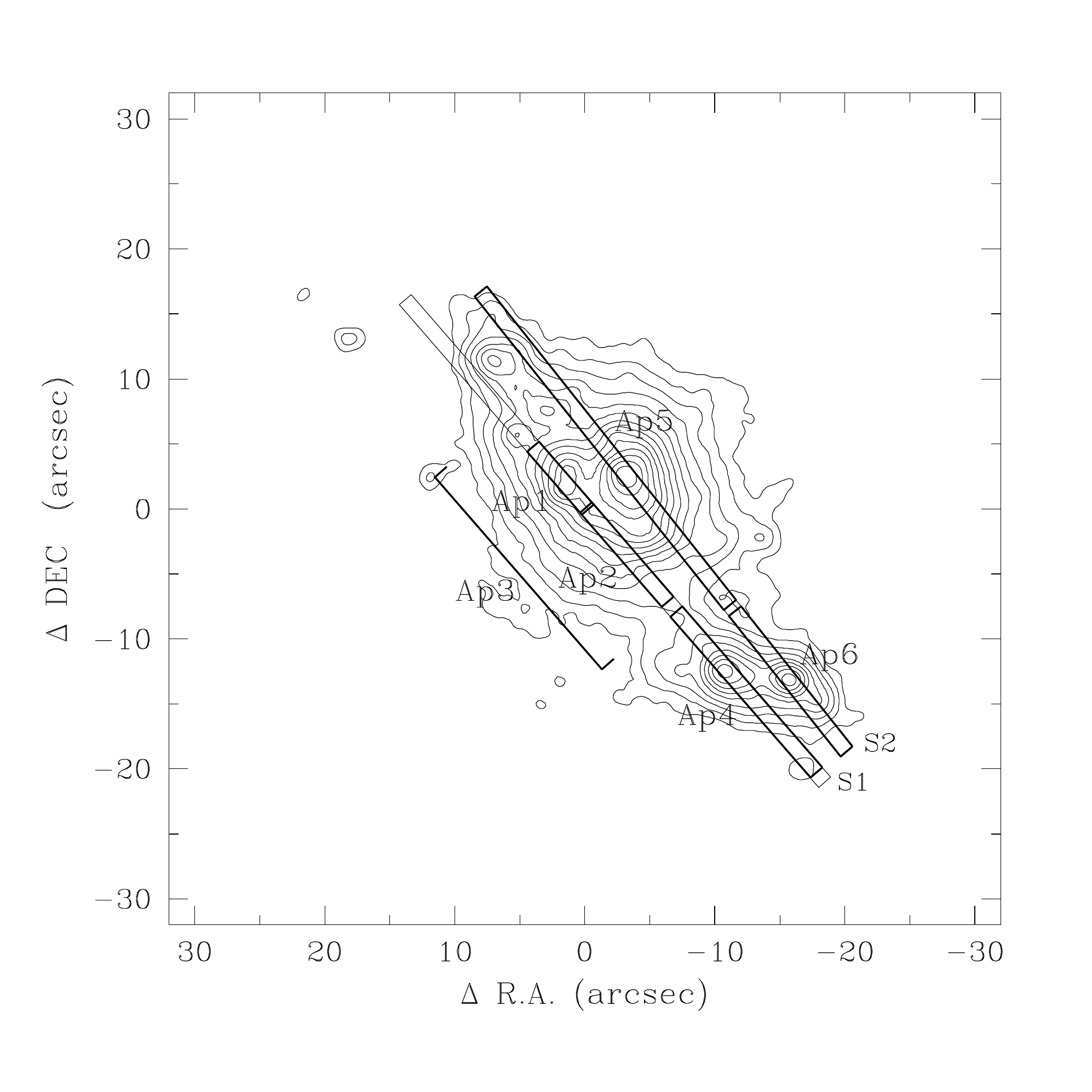}
\includegraphics[width=3.40in]{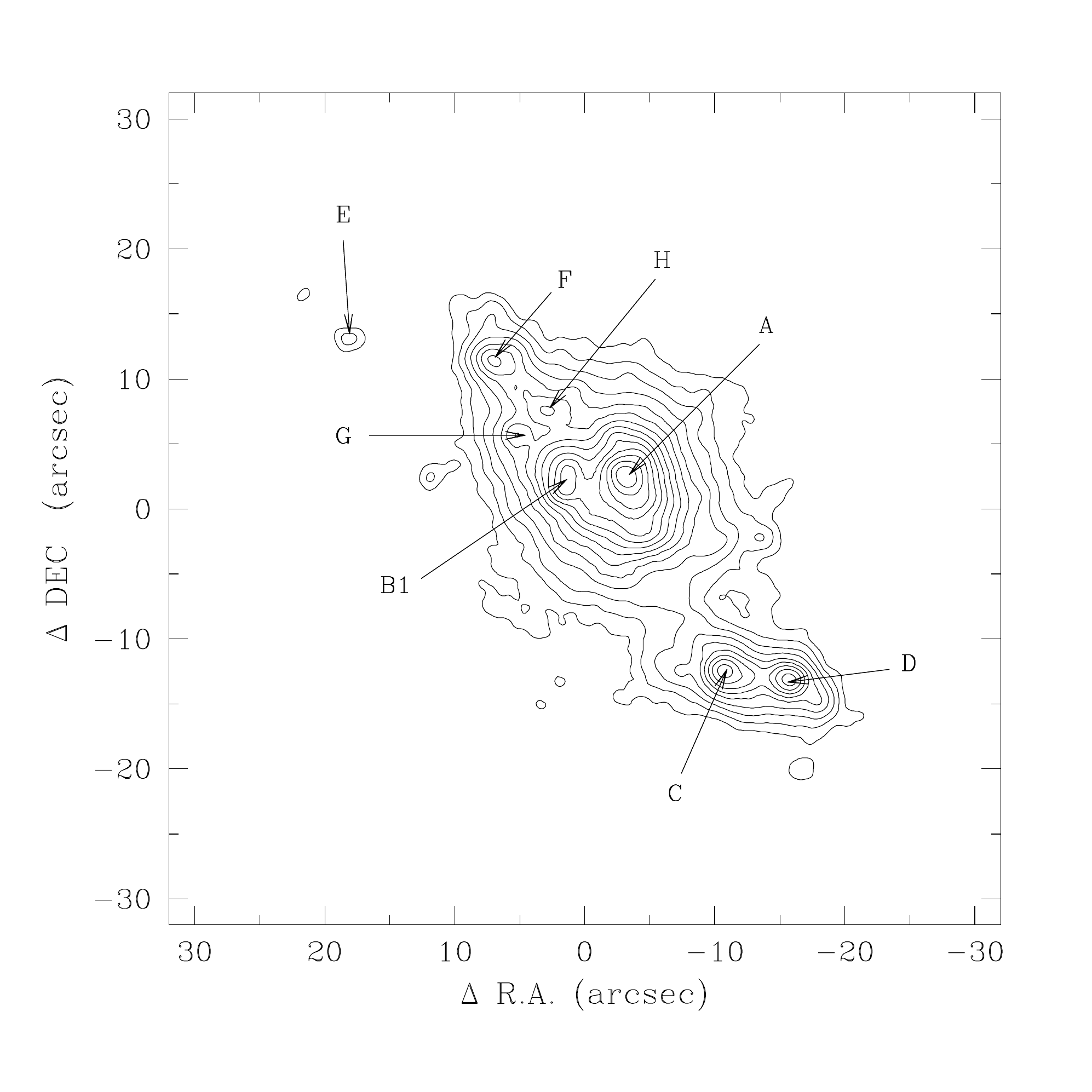}
}}
\vspace*{10pt}
\centerline{\hbox{
\includegraphics[bb=25 36 565 560,width=3.15in,clip=true]{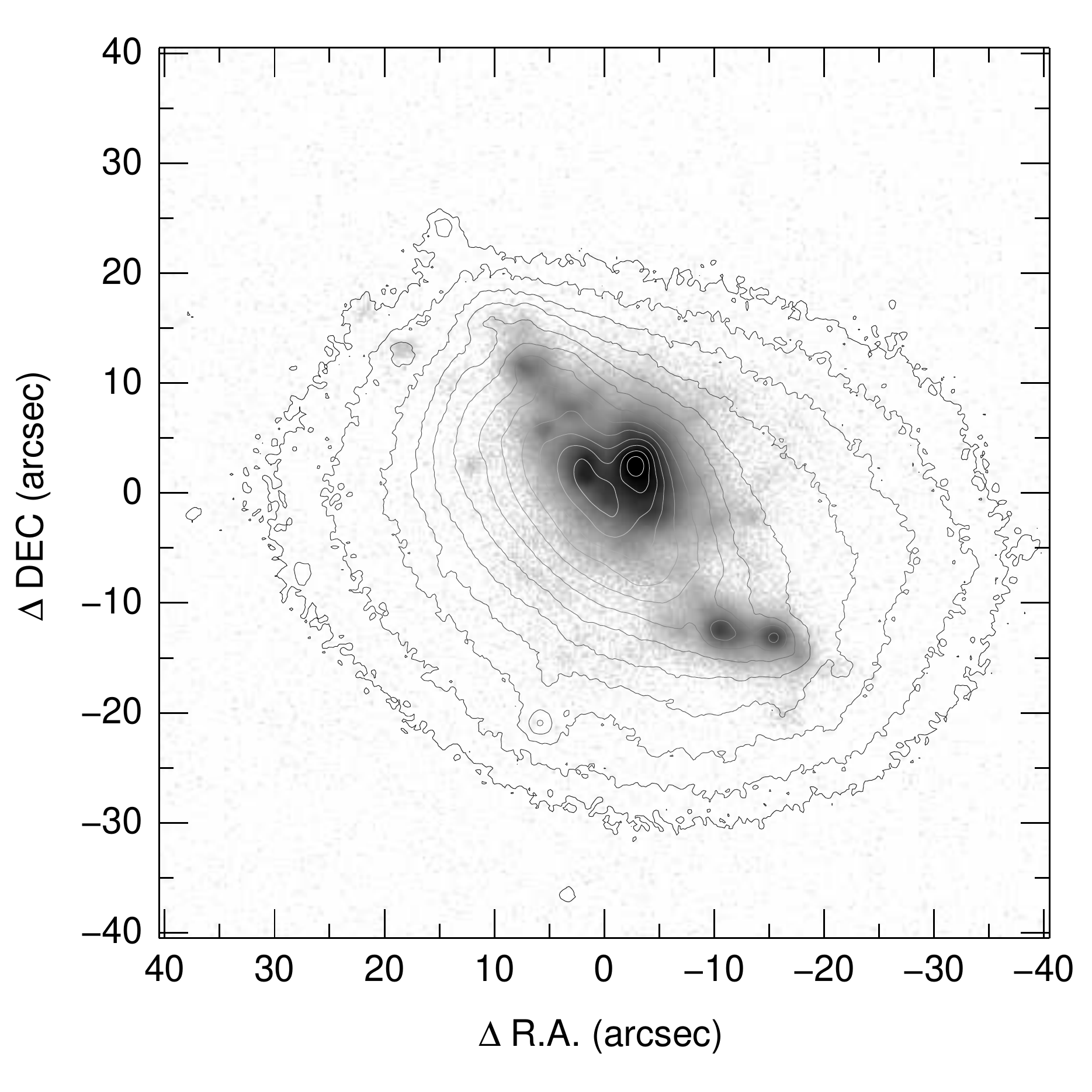}
\hspace*{12pt}
\includegraphics[bb=25 36 565 560,width=3.15in,clip=true]{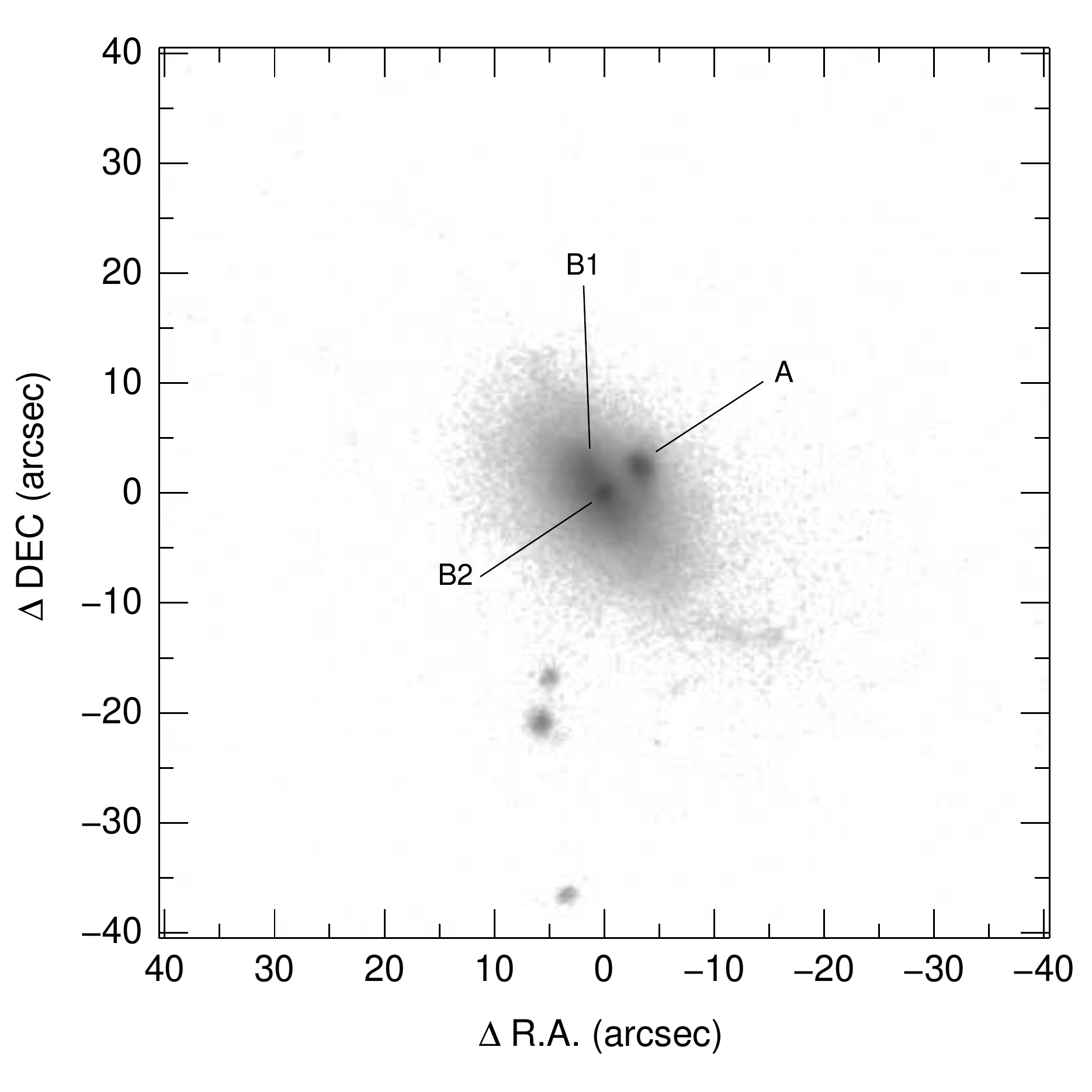}
}}
\caption{
\textit{a) top-left}: Contour plot of the continuum subtracted H$\alpha$ image 
of Mrk~35. The two slit positions are plotted, and the different subregions 
we selected from the spectra are marked in bold line.\
\textit{b) top-right}: Contour plot of the continuum subtracted H$\alpha$ image 
of Mrk~35 with labels identifying the individual star-forming knots.\
\textit{c) bottom-left}: $B$-band contours overlaid on the \Ha\ gray-scale 
map.\ 
\textit{d) bottom-right}: $K$-band map. Knot \knotB2\ is the optical center of 
the galaxy, but is not a \Ha\ peak.
Image orientation is north up and east left. 
Axis units are arcseconds.}
\label{Fig:slitpos}
\end{figure*}

Spectra of Mrk~35 were taken in two almost parallel positions (P.A. $42\degr$
and $38\degr$), as shown in Fig.~\ref{Fig:slitpos}, which pass through the  five
brightest SF knots of the galaxy.  The spatial distribution of the intensity of
the brightest emission lines  (\Ha, H$\beta$, [\ion{O}{3}]) has been used as a
base for defining the regions  from where one-dimensional spectra of each region
were extracted.  The different spatial regions we defined are shown in
Fig.~\ref{Fig:slitpos}: the spatial length of the slit in Position~1 (S1) was 
divided into three subregions, while Position~2 (S2) was split into two
subregions.

\begin{figure*}   
\vspace*{-5mm}
\includegraphics[angle=0,width=\textwidth]{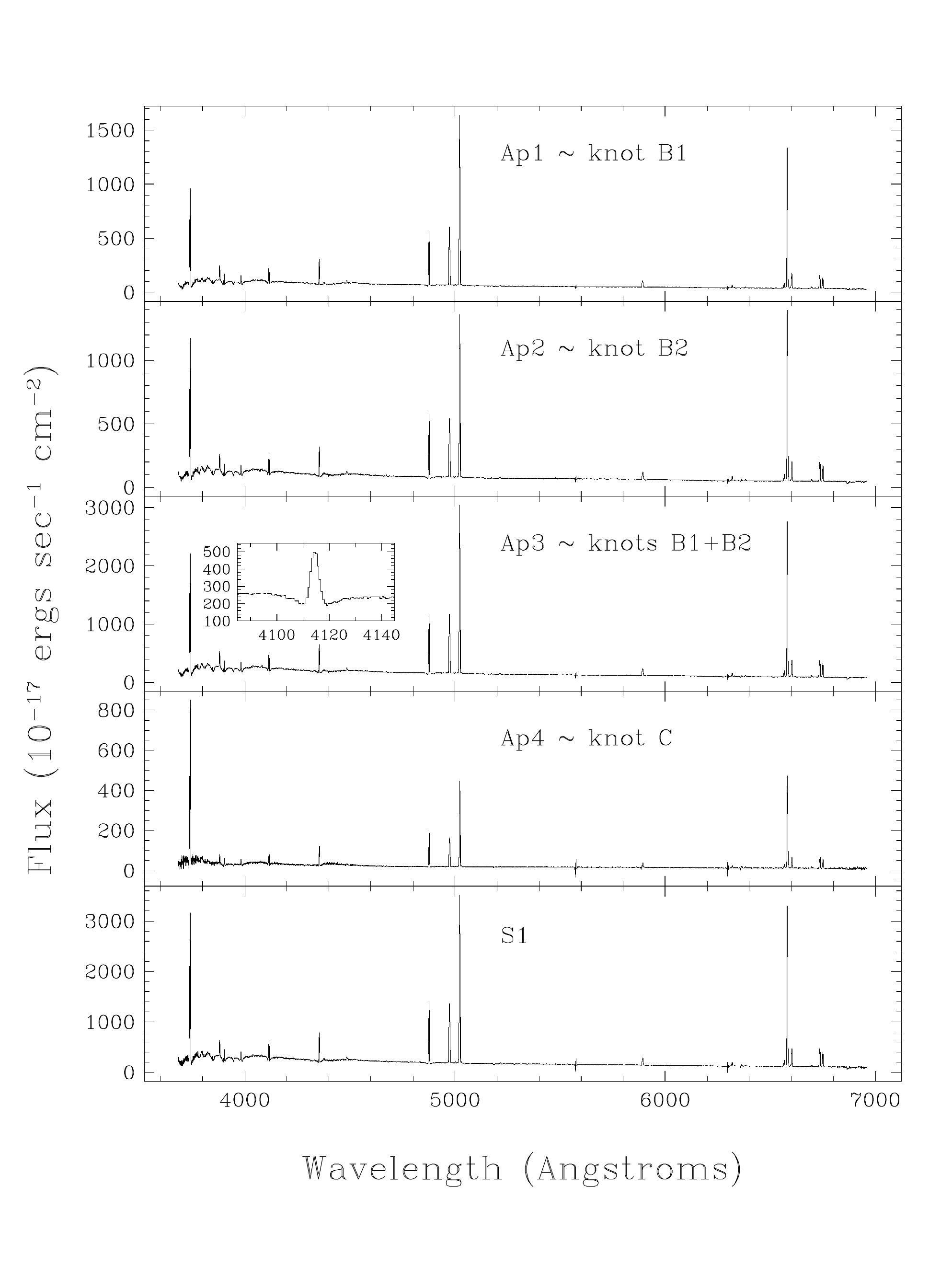}
\caption{Spectra of the different subregions defined in slit position 
S1, and its total integrated spectrum \Sone. The label indicates which knot 
each subregion includes. The inset is an enlargement of the spectrum around 
the \Hd\ line, to enhance the visibility of its absorption wings.}
\label{Fig:spectra1}
\end{figure*}

\begin{figure*}   
\includegraphics[angle=0,width=\textwidth]{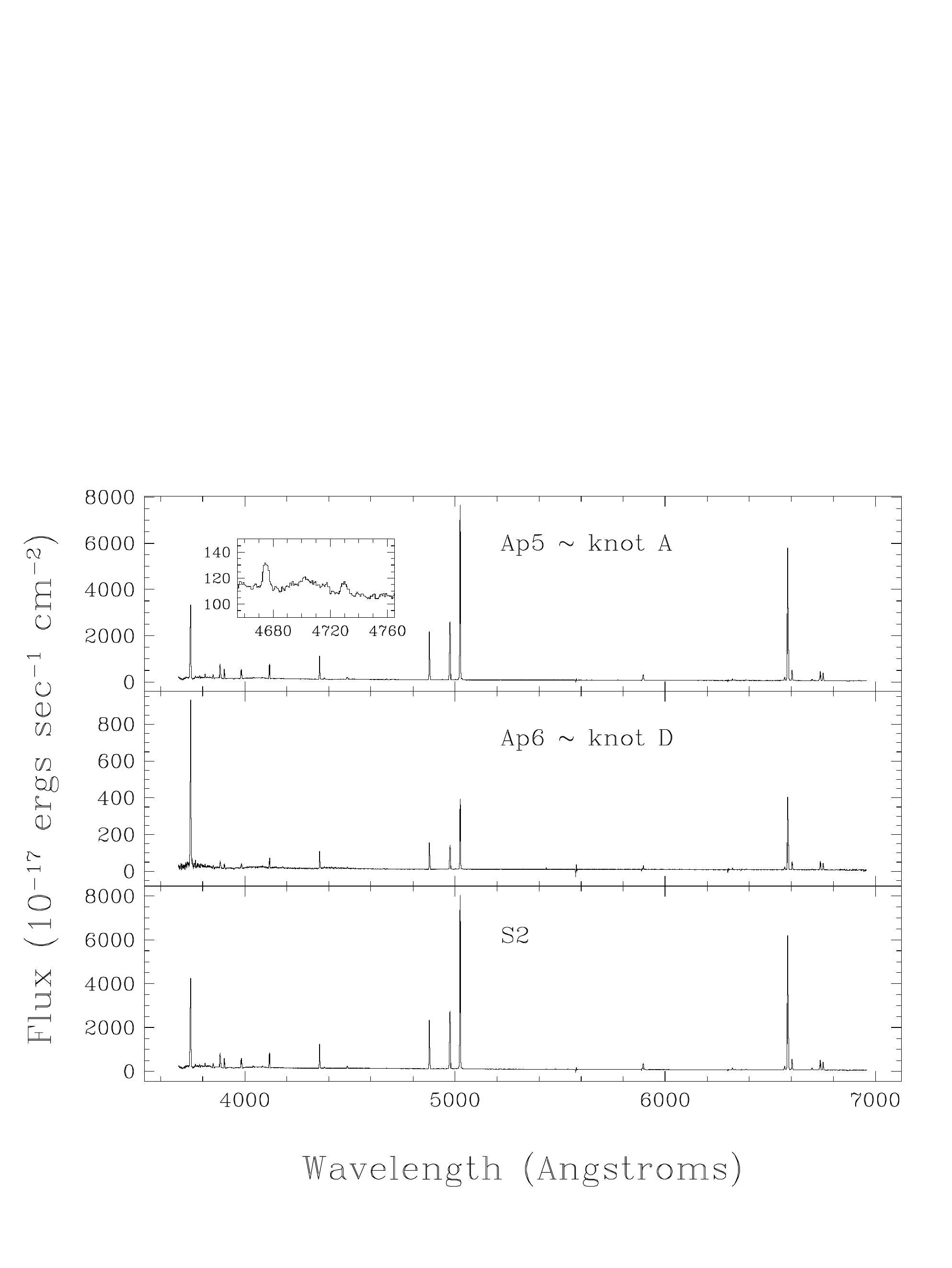}
\caption{Spectra of the different regions defined in slit position S2,
and the total integrated spectrum \Stwo. The inset shows the blue Wolf-Rayet 
bump.}
\label{Fig:spectra2}
\end{figure*}

Figs.~\ref{Fig:spectra1} and \ref{Fig:spectra2} display the integrated spectra
of the different spatial regions, together with the spectrum integrated over 
the whole slit length for \Sone\ and \Stwo. We summarize below the main 
characteristics of the spectra.

\begin{itemize} 

\item \Apone\ includes knot \knotBone, a markedly red SF region ($\BV=0.64$;
$\VI=0.53$) located very close to the optical center of the galaxy (see the 
\Ha\ map in Fig.~\ref{Fig:slitpos}b).  Absorption wings underneath the
Balmer emission lines are clearly visible in \Hb, \Hg\ and \Hd; other
absorption features such as \ion{Ca}{2}~$\lambda$3933, the \ion{Mg}{1}$\,b$
triplet $\lambda\lambda$5167, 5173, 5184 and \ion{Fe}{1}~$\lambda$5335 are
also visible, which indicates the presence of an evolved population of stars
($>10$ Myr). 

\item \Aptwo\ contains knot \knotBtwo, the optical center of the galaxy, which
also is a rather red knot ($\BV=0.63$; $\VI=0.70$; see 
Fig.~\ref{Fig:slitpos}); interestingly, this source is a strong  emitter in
the broad-band frames ---~in the NIR it is, indeed, the  emission peak~--- 
but is not an \Ha\ peak. This implies that \knotBtwo\ is substantially  more
evolved. The \ion{Ca}{2}~$\lambda$3933, the \ion{Mg}{1}$\,b$ triplet and
\ion{Fe}{1}~$\lambda$5335  absorption features are visible.

\item \Apthree\ is the sum of the \Apone\ and \Aptwo.

\item \Apfour\ corresponds to knot \knotC, a SF region located in the 
galaxy "tail-like" feature. It shows a flat spectrum, with prominent emission 
lines and no absorption features, which is characteristic of a dominant OB 
population.

\item \Apfive\ corresponds to knot \knotA, the brightest SF region of the 
galaxy. It shows a flat spectrum characterized by strong nebular emission 
lines, and no visible absorption features. The Wolf-Rayet bump is clearly 
discernible, as already reported by \cite{Steel96} and \cite{Huang99}.

\item \Apsix\ corresponds to knot \knotD, also detached from the main body 
of the galaxy. It too has a flat spectrum, without absorption features.
\end{itemize}

\subsubsection{Reddening Corrected Line Fluxes}

Fluxes and equivalent widths of the emission lines were measured using the
Gaussian profile fitting option in the iraf task {\sc splot} (a direct
integration of the flux for each line gave virtually identical results).

It is known that these measurements are an underestimate of the real flux
of the Balmer lines, because of the underlying absorption component.
To correct for the underlying stellar absorption, some authors 
\citep{Mccall85,Skillman93} adopt a constant equivalent width ($1.5-2$ \AA) 
for all of the hydrogen absorption lines. However, the actual value of the
absorption line equivalent width is uncertain, as it depends on the age of 
the star formation burst \citep{Diaz88,Cananzi93,Olofsson95}, and on 
the contribution of the underlying population of older stars.

The correction from underlying stellar absorption was done in two
different ways, as explained below.

For the Balmer lines showing clear absorption wings, we fitted simultaneously
an absorption and an emission component, using the deblending option in 
{\sc splot}.

In the other cases, we proceeded as follows. We first adopt an initial 
estimate for the absorption equivalent width, $W_{\rm abs}$, correct the 
measured fluxes, and determine the extinction coefficient at 
$\lambda=4861$ \AA, $C(\Hb)$, through a least-square fit to the Balmer 
decrement given by the equation:

\begin{equation}
\label{redde}
\frac{F(\lambda)}{F(\Hb)}=\frac{I(\lambda)}{I(\Hb)} 
\times 10^{C({\rm H}\beta)\times f(\lambda)} 
\end{equation}

\noindent
where $\frac{F(\lambda)}{F({\rm H}\beta)}$ is the line flux corrected for 
absorption and normalized to \Hb;  $\frac{I(\lambda)}{I({\rm H}\beta)}$ is the 
theoretical value for case B recombination, from \citet{Brocklehurst71}, and 
$f(\lambda)$ is the reddening curve normalized to \Hb\ which we took from 
\citet{Whitford58}.

We then vary the value of $W_{\rm abs}$, until we find the one that provides 
the best match (e.g. the minimum scatter in the above relation) between the 
corrected and the theoretical line ratios.
This is done separately for each knot.

Reddening-corrected intensity ratios and equivalent widths for the different
spatial regions, in the two observed positions, are listed in 
Tables~\ref{Tab:fluxpos1a} and \ref{Tab:fluxpos2}.

\begin{deluxetable*}{cccccccc}
\tabletypesize{\tiny}
\tablewidth{0pt}
\tablecaption{Reddening corrected line intensity ratios}
\tablehead{\colhead{Line}  &  \colhead{Ion}     & 
\multicolumn{2}{c}{\Apone} & \multicolumn{2}{c}{\Aptwo}  
& \multicolumn{2}{c}{\Apthree} \\
\colhead{$(\AA)$}            &                        & 
\colhead{$F_{\lambda}$}   & \colhead{$-W_{\lambda}$}  & 
\colhead{$F_{\lambda}$}   & \colhead{$-W_{\lambda}$}  & 
\colhead{$F_{\lambda}$}   & \colhead{$-W_{\lambda}$}}
\startdata
3727 & [\ion{O}{2}]        & $2.058\pm0.043$ &  $   \phn92.86 \pm12.74$ 
                           & $2.404\pm0.050$ &  $   \phn78.47  \pm8.53$ 
                           & $2.230\pm0.024$ &  $   \phn72.00 \phn2.20$   \\
3798 & H10                 & $0.046\pm0.037$ &  $\phn\phn1.14\pm\phn0.95$ 
                           &   \nodata       &    \nodata          
                           &   \nodata       &    \nodata                 \\
3835 & H9                  & $0.059\pm0.019$ &  $\phn\phn1.80\pm\phn0.72$ 
                           & $0.044\pm0.012$ &  $\phn\phn1.08\pm0.35$  
                           & $0.048\pm0.010$ &  $\phn\phn1.25\phn0.30$    \\
3869 & [\ion{Ne}{3}]       & $0.243\pm0.012$ &  $\phn\phn5.38\pm\phn0.38$
                           & $0.216\pm0.012$ &  $\phn\phn3.93\pm0.27$   
                           & $0.227\pm0.005$ &  $\phn\phn4.32\phn0.12$    \\
3889 & \ion{He}{1} + H8    & $0.169\pm0.010$ &  $\phn\phn5.58\pm\phn0.52$
                           & $0.152\pm0.011$ &  $\phn\phn4.19\pm0.42$   
                           & $0.160\pm0.007$ &  $\phn\phn4.62\phn0.28$     \\
3968 & \ion{Ne}{3} + H7    & $0.141\pm0.008$ &  $\phn\phn4.21\pm\phn0.31$
                           & $0.128\pm0.008$ &  $\phn\phn3.20\pm0.25$  
                           & $0.131\pm0.005$ &  $\phn\phn3.44\phn0.17$     \\
4026 & \ion{He}{1}         & $0.031\pm0.018$ &  $\phn\phn0.67\pm\phn0.41$  
                           &   \nodata       &    \nodata         
                           &   \nodata       &    \nodata                  \\
4069 & \ion{S}{2}          & $0.025\pm0.011$ &  $\phn\phn0.55\pm\phn0.25$   
                           &   \nodata       &    \nodata          
                           &   \nodata       &    \nodata                  \\
4101 & H$\delta$           & $0.266\pm0.010$ &  $\phn\phn5.97\pm\phn0.22$ 
                           & $0.297\pm0.011$ &  $\phn\phn5.94\pm0.23$ 
                           & $0.265\pm0.006$ &  $\phn\phn5.17\phn0.13$     \\
4340 & H$\gamma$           & $0.465\pm0.005$ &  $   \phn14.10\pm\phn0.22$    
                           & $0.484\pm0.008$ &  $   \phn11.98\pm0.21$  
                           & $0.468\pm0.005$ &  $   \phn12.38\phn0.16$     \\
4363 & [\ion{O}{3}]        & $0.039\pm0.011$ &  $\phn\phn1.15\pm\phn0.35$    
                           & $0.032\pm0.012$ &  $\phn\phn0.75\pm0.31$  
                           & $0.028\pm0.007$ &  $\phn\phn0.68\phn0.18$     \\
4471 & \ion{He}{1}         & $0.064\pm0.006$ &  $\phn\phn1.77\pm\phn0.20$    
                           &   \nodata       &    \nodata      
                           &   \nodata       &    \nodata                  \\
4861 & \Hb\                & $1.000\pm0.005$ &  $   \phn34.32\pm\phn0.15$   
                           & $1.000\pm0.005$ &  $   \phn29.56\pm0.17$ 
                           & $1.000\pm0.004$ &  $   \phn31.47\phn0.12$     \\
4959 & [\ion{O}{3}]        & $1.025\pm0.005$ &  $   \phn35.08\pm\phn0.22$   
                           & $0.894\pm0.005$ &  $   \phn25.66\pm0.15$ 
                           & $0.936\pm0.004$ &  $   \phn28.16\phn0.14$     \\
5007 & [\ion{O}{3}]        & $3.021\pm0.012$ &  $      103.10\pm\phn0.61$   
                           & $2.540\pm0.011$ &  $   \phn74.90\pm0.31$ 
                           & $2.708\pm0.008$ &  $   \phn81.65\phn0.24$     \\
5200 & [\ion{N}{1}]        & $0.020\pm0.004$ &  $\phn\phn0.80\pm\phn0.17$   
                           & $0.023\pm0.003$ &  $\phn\phn0.78\pm0.12$ 
                           & $0.022\pm0.003$ &  $\phn\phn0.78\phn0.10$     \\
5876 & \ion{He}{1}         & $0.127\pm0.003$ &  $\phn\phn6.19\pm\phn0.22$   
                           & $0.118\pm0.003$ &  $\phn\phn4.63\pm0.14$ 
                           & $0.123\pm0.002$ &  $\phn\phn5.23\phn0.11$     \\
6300 & \ion{O}{1}          & $0.046\pm0.004$ &  $\phn\phn2.65\pm\phn0.26$   
                           & $0.061\pm0.004$ &  $\phn\phn2.85\pm0.20$
                           & $0.054\pm0.002$ &  $\phn\phn2.66\phn0.11$     \\
6312 & \ion{S}{3}          & $0.016\pm0.003$ &  $\phn\phn0.91\pm\phn0.20$   
                           &   \nodata       &    \nodata       
                           & $0.014\pm0.002$ &  $\phn\phn0.68\phn0.13$     \\
6363 & \ion{O}{1}          & $0.013\pm0.002$ &  $\phn\phn0.75\pm\phn0.14$    
                           & $0.021\pm0.004$ &  $\phn\phn1.00\pm0.20$
                           & $0.016\pm0.002$ &  $\phn\phn0.78\phn0.09$     \\
6548 & [\ion{N}{2}]        & $0.092\pm0.003$ &  $\phn\phn5.37\pm\phn0.28$   
                           & $0.110\pm0.003$ &  $\phn\phn5.26\pm0.18$
                           & $0.101\pm0.002$ &  $\phn\phn5.21\phn0.13$     \\
6563 & \Ha\                & $2.856\pm0.017$ &  $      165.40\pm\phn1.96$   
                           & $2.874\pm0.038$ &  $      140.40\pm1.16$
                           & $2.862\pm0.012$ &  $      146.80\phn0.85$     \\
6584 & [\ion{N}{2}]        & $0.292\pm0.003$ &  $   \phn17.30\pm\phn0.34$   
                           & $0.325\pm0.005$ &  $   \phn15.78\pm0.26$ 
                           & $0.308\pm0.003$ &  $   \phn16.19\phn0.20$     \\
6678 & \ion{He}{1}         & $0.032\pm0.005$ &  $\phn\phn2.05\pm\phn0.36$   
                           & $0.037\pm0.005$ &  $\phn\phn1.88\pm0.27$ 
                           & $0.032\pm0.003$ &  $\phn\phn1.75\phn0.14$     \\
6717& [\ion{S}{2}]         & $0.264\pm0.004$ &  $   \phn16.91\pm\phn0.37$    
                           & $0.341\pm0.006$ &  $   \phn17.20\pm0.32$
                           & $0.303\pm0.002$ &  $   \phn16.48\phn0.18$     \\
6731& [\ion{S}{2}]         & $0.208\pm0.004$ &  $   \phn12.94\pm\phn0.34$    
                           & $0.261\pm0.005$ &  $   \phn13.33\pm0.30$
                           & $0.234\pm0.002$ &  $   \phn12.73\phn0.14$     \\[6pt]
$C(\Hb)$   &         &  
\multicolumn{2}{c}{$0.050\pm0.005$} & \multicolumn{2}{c}{$0.064\pm0.016$} &
\multicolumn{2}{c}{$0.067\pm0.004$}  \\
W$(\Ha)_{\rm abs} (\AA)$ &   &\multicolumn{2}{c}{6.0} & \multicolumn{2}{c}{0.0}  & \multicolumn{2}{c}{5.5}  \\
W$(\Hb)_{\rm abs} (\AA)$ &   &\multicolumn{2}{c}{3.8} & \multicolumn{2}{c}{4.2}  & \multicolumn{2}{c}{3.5}  \\
W$(\Hg)_{\rm abs} (\AA)$ &   &\multicolumn{2}{c}{2.9} & \multicolumn{2}{c}{3.5}  & \multicolumn{2}{c}{3.0}  \\
W$(\Hd)_{\rm abs} (\AA)$ &   &\multicolumn{2}{c}{4.2} & \multicolumn{2}{c}{3.5}  & \multicolumn{2}{c}{4.5}  \\
$F(\Hb)$                 &   & \multicolumn{2}{c}{$25.2\pm0.4$}  &
\multicolumn{2}{c}{$27.1\pm1.1$}  & \multicolumn{2}{c}{$55.6\pm0.6$} \\[8pt] 
\hline\\[-6pt]\hline\\[-6pt]
\colhead{{Line}}  &  \colhead{{Ion}}     & 
\multicolumn{2}{c}{\Apfour} &  \multicolumn{2}{c}{\Sone}   & \multicolumn{2}{c}{} \\
\colhead{{$(\AA)$}}            &                & 
\colhead{{$F_{\lambda}$}}   & \colhead{{$-W_{\lambda}$}}  & 
\colhead{{$F_{\lambda}$}}   & \colhead{{$-W_{\lambda}$}}  & &  \\[4pt]
\hline\\[-4pt]
3727 & [\ion{O}{2}]        &   \nodata       &    \nodata	    
                           &   \nodata       &    \nodata	  
			   &                 &                            \\
3798 & H10                 &   \nodata       &    \nodata             
                           &   \nodata       &    \nodata         
			   &                 &                            \\
3835 & H9                  &   \nodata       &    \nodata             
                           & $0.039\pm0.007$ &  $\phn\phn1.04\pm0.21$ 
			   &                 &                            \\
3869 & [\ion{Ne}{3}]       & $0.212\pm0.020$ &  $\phn\phn4.81\pm\phn0.64$ 
                           & $0.203\pm0.005$ &  $\phn\phn4.06\pm0.12$ 
			   &                 &                            \\
3889 & \ion{He}{1} + H8    & $0.188\pm0.021$ &  $\phn\phn6.13\pm\phn0.97$ 
                           & $0.145\pm0.005$ &  $\phn\phn4.31\pm0.18$ 
			   &                 &                            \\
3968 & \ion{Ne}{3} + H7    & $0.166\pm0.019$ &  $\phn\phn4.68\pm\phn0.66$
                           & $0.125\pm0.004$ &  $\phn\phn3.48\pm0.12$ 
			   &                 &                            \\
4026 & \ion{He}{1}         &   \nodata       &    \nodata     
                           &   \nodata       &    \nodata                
			   &                 &                            \\
4069 & \ion{S}{2}          &   \nodata       &    \nodata        
                           &   \nodata       &    \nodata                
			   &                 &                             \\
4101 & H$\delta$           & $0.343\pm0.024$ &  $\phn\phn6.81\pm\phn0.50$ 
                           & $0.262\pm0.004$ &  $\phn\phn5.30\pm0.09$
			   &                 &                             \\
4340 & H$\gamma$           & $0.524\pm0.018$ &  $   \phn15.65\pm\phn0.79$  
                           & $0.471\pm0.003$ &  $   \phn13.41\pm0.12$ 
			   &                 &                             \\
4363 & [\ion{O}{3}]        &   \nodata       &    \nodata     
                           & $0.019\pm0.005$ &  $\phn\phn0.47\pm0.13$ 
			   &                 &                             \\
4471 & \ion{He}{1}         & $0.037\pm0.012$ &  $\phn\phn0.95\pm\phn0.35$   
                           & $0.049\pm0.004$ &  $\phn\phn1.19\pm0.11$ 
			   &                 &                             \\
4861 & \Hb\                & $1.000\pm0.011$ &  $   \phn40.34\pm\phn0.74$
                           & $1.000\pm0.003$ &  $   \phn34.62\pm0.24$ 
			   &                 &                             \\
4959 & [\ion{O}{3}]        & $0.823\pm0.009$ &  $   \phn29.79\pm\phn0.37$
                           & $0.878\pm0.003$ &  $   \phn26.88\pm0.10$ 
			   &                 &                             \\
5007 & [\ion{O}{3}]        & $2.399\pm0.022$ &  $   \phn85.79\pm\phn0.84$
                           & $2.552\pm0.007$ &  $   \phn79.29\pm0.26$ 
			   &                 &                             \\
5200 & [\ion{N}{1}]        & $0.034\pm0.010$ &  $\phn\phn1.44\pm\phn0.44$ 
                           & $0.024\pm0.002$ &  $\phn\phn0.86\pm0.08$  
			   &                 &                             \\
5876 & \ion{He}{1}         & $0.119\pm0.007$ &  $\phn\phn6.06\pm\phn0.55$
                           & $0.121\pm0.002$ &  $\phn\phn5.02\pm0.07$  
			   &                 &                             \\
6300 & \ion{O}{1}          & $0.078\pm0.010$ &  $\phn\phn4.44\pm\phn0.73$ 
                           & $0.058\pm0.001$ &  $\phn\phn2.77\pm0.08$  
			   &                 &                             \\
6312 & \ion{S}{3}          &   \nodata       &    \nodata    
                           & $0.015\pm0.002$ &  $\phn\phn0.75\pm0.12$  
			   &                 &                             \\
6363 & \ion{O}{1}          & $0.031\pm0.010$ &  $\phn\phn1.76\pm\phn0.63$
                           & $0.018\pm0.001$ &  $\phn\phn0.87\pm0.07$ 
			   &                 &                             \\
6548 & [\ion{N}{2}]        & $0.111\pm0.009$ &  $\phn\phn6.18\pm\phn0.50$  
                           & $0.104\pm0.002$ &  $\phn\phn5.09\pm0.12$ 
			   &                 &                             \\
6563 & \Ha\                & $2.925\pm0.140$ &  $      162.90\pm\phn3.04$ 
                           & $2.866\pm0.011$ &  $      141.30\pm0.73$ 
			   &                 &                             \\
6584 & [\ion{N}{2}]        & $0.327\pm0.017$ &  $   \phn17.99\pm\phn0.57$
                           & $0.314\pm0.002$ &  $   \phn15.61\pm0.16$ 
			   &                 &                             \\
6678 & \ion{He}{1}         & $0.028\pm0.007$ &  $\phn\phn1.66\pm\phn0.50$ 
                           & $0.031\pm0.002$ &  $\phn\phn1.62\pm0.09$ 
			   &                 &                             \\
6717 &[\ion{S}{2}]         & $0.376\pm0.020$ &  $   \phn22.97\pm\phn0.78$ 
                           & $0.325\pm0.002$ &  $   \phn16.97\pm0.20$ 
			   &                 &                            \\
6731 &[\ion{S}{2}]         & $0.277\pm0.015$ &  $   \phn18.04\pm\phn0.71$
                           & $0.245\pm0.002$ &  $   \phn12.76\pm0.17$
			   &                 &                    	   \\[6pt]
$C(\Hb)$   &         &  
\multicolumn{2}{c}{$0.035\pm0.060$} &
\multicolumn{2}{c}{$0.001\pm0.003$} \\
W$(\Ha)_{\rm abs} (\AA)$ &     & \multicolumn{2}{c}{0.0}  & \multicolumn{2}{c}{2.5} & \multicolumn{2}{c}{} \\
W$(\Hb)_{\rm abs} (\AA)$ &     & \multicolumn{2}{c}{0.0}  & \multicolumn{2}{c}{2.5} & \multicolumn{2}{c}{} \\
W$(\Hg)_{\rm abs} (\AA)$ &     & \multicolumn{2}{c}{0.0}  & \multicolumn{2}{c}{2.3} & \multicolumn{2}{c}{} \\
W$(\Hd)_{\rm abs} (\AA)$ &     & \multicolumn{2}{c}{5.0}  & \multicolumn{2}{c}{4.5} & \multicolumn{2}{c}{} \\
$F(\Hb)$    &          & 
\multicolumn{2}{c}{$8.1\pm1.2$} & \multicolumn{2}{c}{$57.8\pm0.5$}  & \multicolumn{2}{c}{} \\  
    
\enddata
\tabletypesize{\footnotesize}
\tablecomments{Reddening-corrected line intensities, normalized to $\Hb=1$, 
for the measured apertures from the long slit spectrum in position 1.
Balmer lines are corrected from underlying stellar absorption.
The reddening coefficient, $C(\Hb)$, the value of the absorption correction 
in the Balmer lines, W$_{\rm abs}$, and the corrected \Hb\ flux, $F(\Hb)$ 
($\times 10^{-15}$ \ergscms), are also listed.
[\ion{O}{2}] data for
Ap4 and S1 are affected by "ghost emission" from knot \knotBone\ and are 
omitted (see text in \S~\ref{Section:PhysicalConditions} for details).
\label{Tab:fluxpos1a} }
\end{deluxetable*}

\begin{deluxetable*}{cccccccc}
\tabletypesize{\tiny}
\tablewidth{0pt}
\tablecaption{Reddening corrected line intensity ratios (Position 2)}
\tablehead{\colhead{Line}  &  \colhead{Ion}     & 
\multicolumn{2}{c}{\Apfive} & \multicolumn{2}{c}{\Apsix}  &
\multicolumn{2}{c}{\Stwo} \\
\colhead{$(\AA)$}            &                        & 
\colhead{$F_{\lambda}$}   & \colhead{$-W_{\lambda}$}  & 
\colhead{$F_{\lambda}$}   & \colhead{$-W_{\lambda}$}  &  
\colhead{$F_{\lambda}$}   & \colhead{$-W_{\lambda}$}}
\startdata
3727 & [\ion{O}{2}]      & $2.023\pm0.032$  &  $      117.10\pm4.19$
                         &   \nodata        &    \nodata  
                         &   \nodata        &    \nodata       \\
3798 & H12               &   \nodata        &    \nodata      
                         & $0.165\pm0.119$  &  $\phn\phn5.11\pm\phn3.59$  
                         &   \nodata        &    \nodata                 \\
3798 & H11               & $0.085\pm0.013$  &  $\phn\phn5.52\pm1.43$
                         &   \nodata        &    \nodata      
                         & $0.059\pm0.012$  &  $\phn\phn3.05\pm0.80$ \\
3798 & H10               & $0.071\pm0.008$  &  $\phn\phn3.33\pm0.48$
                         &   \nodata        &    \nodata      
                         & $0.071\pm0.007$  &  $\phn\phn3.24\pm0.47$ \\
3835 & H9                & $0.081\pm0.004$  &  $\phn\phn4.72\pm0.36$
                         & $0.121\pm0.052$  &  $\phn\phn5.33\pm\phn3.27$
                         & $0.080\pm0.004$  &  $\phn\phn4.53\pm0.31$ \\
3869 & [\ion{Ne}{3}]     & $0.274\pm0.004$  &  $   \phn13.77\pm0.22$
                         & $0.240\pm0.028$  &  $\phn\phn8.21\pm\phn1.45$
                         & $0.267\pm0.005$  &  $   \phn13.23\pm0.17$ \\
3889 & HeI + H8          & $0.190\pm0.003$  &  $   \phn11.88\pm0.33$
                         & $0.189\pm0.028$  &  $\phn\phn8.35\pm\phn1.77$
                         & $0.187\pm0.004$  &  $   \phn11.59\pm0.28$ \\
3968 & NeIII + H7        & $0.229\pm0.004$  &  $   \phn14.10\pm0.28$
                         & $0.197\pm0.022$  &  $\phn\phn7.73\pm\phn1.05$
                         & $0.220\pm0.004$  &  $   \phn13.02\pm0.28$ \\
4026 & HeI               & $0.022\pm0.003$  &  $\phn\phn1.17\pm0.19$
                         & $0.053\pm0.047$  &  $\phn\phn1.57\pm\phn1.58$
                         & $0.023\pm0.003$  &  $\phn\phn1.14\pm0.16$ \\
4069 & \ion{S}{2}        & $0.019\pm0.004$  &  $\phn\phn1.01\pm0.22$
                         &   \nodata        &    \nodata           
                         & $0.018\pm0.003$  &  $\phn\phn0.89\pm0.17$ \\
4101& H$\delta$          & $0.297\pm0.004$  &  $   \phn17.83\pm0.34$
                         & $0.365\pm0.041$  &  $   \phn13.94\pm\phn3.18$    
                         & $0.299\pm0.005$  &  $   \phn17.28\pm0.35$ \\
4340& H$\gamma$          & $0.489\pm0.004$  &  $   \phn36.25\pm0.50$
                         & $0.594\pm0.028$  &  $   \phn23.10\pm\phn1.29$
                         & $0.494\pm0.005$  &  $   \phn34.67\pm0.43$ \\
4363& [\ion{O}{3}]       & $0.025\pm0.002$  &  $\phn\phn1.65\pm0.12$
                         & $0.064\pm0.031$  &  $\phn\phn2.25\pm\phn1.30$
                         & $0.030\pm0.002$  &  $\phn\phn1.93\pm0.15$ \\
4471 & \ion{He}{1}       & $0.045\pm0.001$  &  $\phn\phn3.05\pm0.12$     
                         & $0.046\pm0.012$  &  $\phn\phn1.82\pm\phn0.58$
                         & $0.045\pm0.002$  &  $\phn\phn2.95\pm0.11$ \\
4861 & \Hb\              & $1.000\pm0.002$  &  $   \phn90.83\pm0.40$     
                         & $1.000\pm0.010$  &  $   \phn54.18\pm\phn0.84$
                         & $1.000\pm0.002$  &  $   \phn87.46\pm0.32$ \\
4959 & [\ion{O}{3}]      & $1.222\pm0.003$  &  $      102.50\pm0.36$
                         & $0.890\pm0.011$  &  $   \phn42.73\pm\phn0.83$
                         & $1.192\pm0.003$  &  $   \phn96.39\pm0.31$ \\
5007 & [\ion{O}{3}]      & $3.597\pm0.009$  &  $      272.40\pm0.92$     
                         & $2.603\pm0.032$  &  $      120.10\pm\phn1.68$
                         & $3.506\pm0.010$  &  $      247.70\pm0.82$ \\
5200 & [\ion{N}{1}]      & $0.013\pm0.001$  &  $\phn\phn1.35\pm0.12$     
                         & $0.030\pm0.008$  &  $\phn\phn1.72\pm\phn0.53$
                         & $0.013\pm0.001$  &  $\phn\phn1.35\pm0.13$ \\
5411 & \ion{He}{2}       &   \nodata        &    \nodata           
                         &   \nodata        &    \nodata            
                         &   \nodata        &    \nodata                 \\
5876 & \ion{He}{1}       & $0.128\pm0.002$  &  $   \phn16.44\pm0.26$    
                         & $0.119\pm0.010$  &  $\phn\phn7.80\pm\phn0.56$
                         & $0.126\pm0.002$  &  $   \phn14.84\pm0.22$ \\
6300 & \ion{O}{1}        & $0.037\pm0.001$  &  $\phn\phn5.78\pm0.24$     
                         & $0.068\pm0.010$  &  $\phn\phn5.25\pm\phn0.78$
                         & $0.039\pm0.001$  &  $\phn\phn5.62\pm0.21$ \\
6312 & \ion{S}{3}        & $0.014\pm0.001$  &  $\phn\phn2.06\pm0.14$
                         &   \nodata        &    \nodata            
                         & $0.013\pm0.001$  &  $\phn\phn1.85\pm0.13$ \\
6363 & \ion{O}{1}        & $0.012\pm0.001$  &  $\phn\phn1.82\pm0.15$  
                         & $0.023\pm0.007$  &  $\phn\phn1.74\pm\phn0.55$
                         & $0.013\pm0.001$  &  $\phn\phn1.78\pm0.12$ \\
6548 & [\ion{N}{2}]      & $0.071\pm0.002$  &  $\phn\phn9.28\pm0.22$     
                         & $0.102\pm0.012$  &  $\phn\phn6.84\pm\phn0.62$
                         & $0.078\pm0.002$  &  $   \phn10.26\pm0.23$ \\
6563 & \Ha\              & $2.885\pm0.053$  &  $      344.00\pm2.58$    
                         & $2.998\pm0.301$  &  $      220.80\pm\phn8.17$
                         & $2.900\pm0.072$  &  $      306.40\pm1.57$ \\
6584 & [\ion{N}{2}]      & $0.221\pm0.004$  &  $   \phn31.15\pm0.39$     
                         & $0.317\pm0.033$  &  $   \phn22.84\pm\phn0.98$
                         & $0.229\pm0.006$  &  $   \phn30.01\pm0.34$ \\
6678 & \ion{He}{1}       & $0.034\pm0.001$  &  $\phn\phn5.66\pm0.19$ 
                         & $0.041\pm0.010$  &  $\phn\phn3.51\pm\phn1.07$
                         & $0.035\pm0.001$  &  $\phn\phn5.27\pm0.17$ \\
6717 & [\ion{S}{2}]      & $0.201\pm0.004$  &  $   \phn32.07\pm0.45$     
                         & $0.348\pm0.038$  &  $   \phn29.65\pm\phn1.63$
                         & $0.211\pm0.006$  &  $   \phn31.43\pm0.41$ \\
6731 & [\ion{S}{2}]      & $0.160\pm0.003$  &  $   \phn25.88\pm0.33$     
                         & $0.261\pm0.029$  &  $   \phn22.81\pm\phn1.51$
                         & $0.168\pm0.005$  &  $   \phn25.65\pm0.32$ \\[6pt]
$C(\Hb)$     &         & \multicolumn{2}{c}{0.079$\pm$0.023}& 
\multicolumn{2}{c}{0.035$\pm$0.130}&   \multicolumn{2}{c}{0.058$\pm$0.032}\\
W$(\Ha)_{\rm abs} (\AA)$ &       & \multicolumn{2}{c}{0}&\multicolumn{2}{c}{6.0} & \multicolumn{2}{c}{0}\\
W$(\Hb)_{\rm abs} (\AA)$ &       & \multicolumn{2}{c}{0}&\multicolumn{2}{c}{0.0} & \multicolumn{2}{c}{0}\\
W$(\Hg)_{\rm abs} (\AA)$ &       & \multicolumn{2}{c}{0}&\multicolumn{2}{c}{0.0} & \multicolumn{2}{c}{0}\\
W$(\Hd)_{\rm abs} (\AA)$ &       & \multicolumn{2}{c}{0}&\multicolumn{2}{c}{0.0} & \multicolumn{2}{c}{0}\\
$F(\Hb)$    &          & \multicolumn{2}{c}{110.4$\pm$6.7}& \multicolumn{2}{c}{6.8$\pm$2.3} &
\multicolumn{2}{c}{$112.8\pm9.3$}\\      
\enddata
\tabletypesize{\footnotesize}
\tablecomments{Reddening-corrected line intensities, normalized to $\Hb=1$,
for the measured apertures from the long slit spectrum, position 2.
Balmer lines are corrected from underlying stellar absorption.
The reddening coefficient, $C(\Hb)$, the value of the absorption correction 
in the different Balmer lines, W$_{\rm abs}$, and the corrected \Hb\ flux, 
$F(\Hb)$ ($\times 10^{-15}$ \ergscms), are also listed. [\ion{O}{2}] data for
Ap6 and S2 are affected by "ghost emission" from knot \knotA\ and are omitted
(see text in \S~\ref{Section:PhysicalConditions} for details).
\label{Tab:fluxpos2} }
\end{deluxetable*}

\subsubsection{Physical Conditions and Chemical Abundances}
\label{Section:PhysicalConditions}

Table~\ref{Table:abundances} lists the most relevant line ratios, the physical 
parameters and oxygen abundances derived for each aperture.

The excitation ratio [\ion{O}{3}]/\Hb{} is $>2$ in all the knots, and it is
highest in knot \knotA, the brightest star-forming region, where it reaches 
$\simeq 3.6$.
[\ion{N}{2}]/\Ha{} drops where [\ion{O}{3}]/\Hb{} increases, as expected in 
regions of recent star-formation.

[\ion{O}{1}]~$\lambda6300$ is detected in all the apertures, but the fluxes 
measured are quite small. 
Photoionization by stars is the dominant mechanism. Although the
presence of shocks cannot be completely ruled out, the small values of
[\ion{S}{2}]/\Ha{} and [\ion{N}{2}]/\Ha{} indicate that shocks do not play a 
significant role. As expected, in the classical diagrams [\ion{N}{2}]/\Ha{} 
vs [\ion{O}{3}]/\Hb, [\ion{S}{2}]/\Ha{} vs
[\ion{O}{3}]\Hb{} and [\ion{O}{1}]/\Ha{} vs [\ion{O}{3}]/\Hb, all the
star-forming knots are located within the locus of \ion{H}{2} regions.
The electron density of the ionized gas was determined from the 
[\ion{S}{2}] $\lambda$6717/$\lambda$6731 ratio. We used the task 
\textsc{temden}, based on the \textsc{fivel} program \citep{ShawDufour95}, 
which is included in the {\sc iraf} package \textsc{nebular}. For knot \knotA\ 
we found a considerably high density (170 cm$^{-3}$), in good agreement with 
the value reported in \cite{IzotovThuan04}. Knots \knotB, \knotC\ and \knotD\ 
have lower gas densities ($\leq 100$ cm$^{-3}$), characteristic of massive 
\ion{H}{2} regions.

A precise determination of the ionized gas temperature (\Te) requires  an
accurate measurement of the flux of [\ion{O}{3}]~$\lambda$4363; though we 
detected this line in five out of the six apertures, the uncertainties in the
flux measurements ---~due to the weakness of the line as well as to the strong
absorption wings of the nearby H$\gamma$ line~--- make the derived values
unreliable. Only in \Apfive\ and \Stwo\ the  uncertainties in the flux 
measurements are less than 25\%, and we could compute \Te\ from the line flux
ratio [\ion{O}{3}] $\lambda$4363/($\lambda4959 + \lambda5007$). 

An alternative method to estimate \Te\ is provided by the empirical relation 
proposed by \cite{Pilyugin01}, which relies on the flux of 
[\ion{O}{2}]~$\lambda3727$.

The [\ion{O}{2}]~$\lambda 3727$ fluxes we measured in knots \knotC\ and \knotD\ 
are abnormally high when compared to their \Hb\ fluxes and give unrealistic 
values for \Te. This problem was found to be caused by strong contamination of
the [\ion{O}{2}] line in knots \knotC\ and \knotD\ by ''ghost [\ion{O}{2}] 
emission'', originating from knots \knotBone\ and \knotA.
This ghost, whose amplitude is about one fourth that of the parent
[\ion{O}{2}] line, is visible in all the spectra of the other objects observed 
in the same run. Oddly enough, no ghost is produced by any other emission 
lines.

Abundances have been estimated from [\ion{O}{2}]~$\lambda3727$ 
\citep{Pilyugin01} and from the empirical N2 calibrator, 
log([\ion{N}{2}] $\lambda$6584/H$\alpha$) introduced by \cite{Storchi94}; 
we used the linear fits derived in  \cite{Denicolo02} and \cite{Pettini04}.

\begin{deluxetable*}{lcccccccc}
\tabletypesize{\footnotesize}
\tablewidth{0pt}
\tablecaption{Physical parameters and oxygen abundances for the apertures
measured in Mrk~35.}
\tablehead{
\colhead{} &
\colhead{\Apone} &
\colhead{\Aptwo} &
\colhead{\Apthree} &
\colhead{\Apfour}  &
\colhead{\Sone} &
\colhead{\Apfive} &
\colhead{\Apsix} &
\colhead{\Stwo} \\
\colhead{} &
\colhead{(knot \knotB1)} &
\colhead{(knot \knotB2)} &
\colhead{(knots \knotB1+\knotB2)} &
\colhead{(knot \knotC)}  &
\colhead{} &
\colhead{(knot \knotA)} &
\colhead{(knot \knotD)} & 
\colhead{} 
}
\startdata
$\log$([\ion{O}{3}]/\Hb)  & \phs0.48& \phs0.40& \phs0.43& \phs0.38& \phs0.41& \phs0.56& \phs0.42& \phs0.54 \\	  
$\log$([\ion{N}{2}]/\Ha)  & $-0.99$ & $-0.95$ & $-0.96$ & $-0.95$ & $-0.96$ & $-1.12$ & $-0.98$ & $-1.10$ \\
$\log$([\ion{S}{2}]/\Ha)  & $-0.78$ & $-0.68$ & $-0.73$ & $-0.65$ & $-0.70$ & $-0.90$ & $-0.69$ & $-0.88$ \\
$\log$([\ion{O}{1}]/\Ha)  & $-1.79$ & $-1.67$ & $-1.72$ & $-1.57$ & $-1.69$ & $-1.89$ & $-1.64$ & $-1.87$\\
 \Ne$\;$(cm$^{-3}$)       & 150     & 110     & 120     & $<100$  & 100     & 170     & $<100$  & 170 \\
$\Te\;(K)$                & \nodata & \nodata & \nodata & \nodata & \nodata & 10250   & \nodata & 10770 \\
$\Te\;(K)$                & 8700    & 8800    & 8700    & \nodata & \nodata & \phn9200& \nodata & \nodata \\
$12+\log$(O/H)            & 8.46    & 8.43    & 8.45    & \nodata & \nodata & 8.43    & \nodata & \nodata \\
$12+\log$(O/H)            & 8.40    & 8.43    & 8.41    & 8.43    & 8.42    & 8.31    & 8.42    & 8.32 \\
$12+\log$(O/H)            & 8.43    & 8.36    & 8.35    & 8.36    & 8.35    & 8.26    & 8.36    & 8.27 \\
\enddata
\label{Table:abundances}
\tablecomments{{}Line 6: \Te\ derived from the 
[\ion{O}{3}] $\lambda$4363/($\lambda4959 + \lambda5007$) ratio; 
line 7: \Te\ derived from \cite{Pilyugin01}; 
line 8: abundance derived from \cite{Pilyugin01}; 
line 9: abundance derived from \cite{Denicolo02}; 
line 10: abundance derived from \cite{Pettini04}.}
\end{deluxetable*}

\subsubsection{Gaseous Kinematics}

The kinematics of the gas was determined by fitting a Gaussian to the
strongest emission lines (H$\alpha$, H$\beta$, [\ion{O}{3}]$\lambda4959$ and
[\ion{O}{3}]$\lambda5007$).
In the regions between or outside the knots where the gas emission is 
fainter, we averaged the spectrum by 3 to 5 pixels in the spatial 
direction to increase the signal-to-noise ratio.
At each spatial position, we averaged together the velocity values of the 
measured emission lines (after rejecting outliers), and took their scatter as 
an estimate of the associated uncertainty.
A zero point offset on the velocity axis was applied to correct for small 
shifts in the wavelength scale, derived by measuring the position of a
few bright sky-lines.

\begin{figure*}   
\vspace*{-7mm}
\includegraphics[angle=270,width=\textwidth]{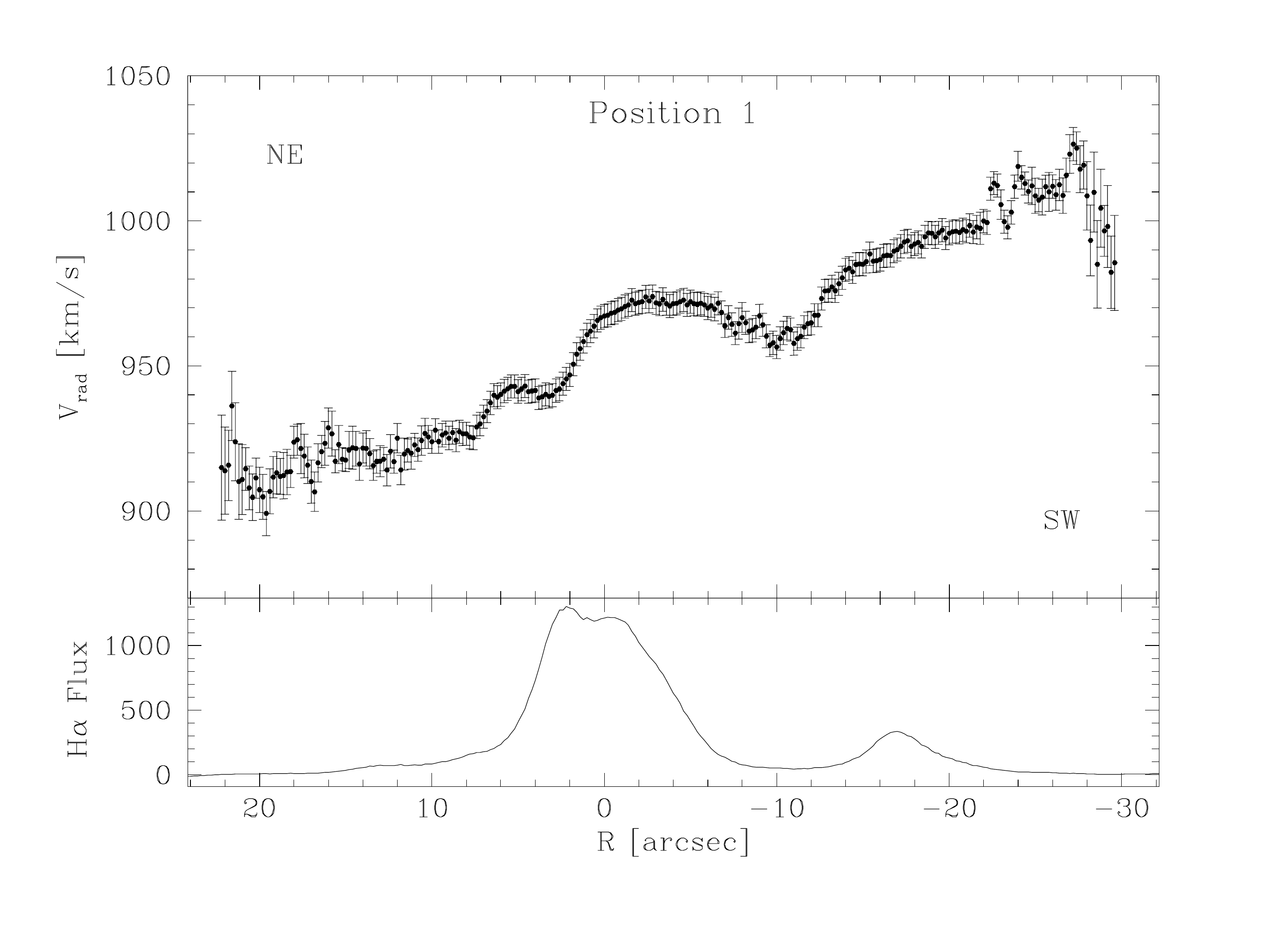}
\caption{Velocity profile for Position 1. The \Ha\ flux profile (in 
$10^{-17}$ \ergscms\ arcsec$^{-2}$ units) is also plotted to help relate 
kinematical and morphological features. Coordinate $R=0$ was set at 
the center of the peak of the continuum emission, \knotBtwo.}
\label{Fig:kin_pos1}
\end{figure*}

\begin{figure*}   
\includegraphics[angle=270,width=\textwidth]{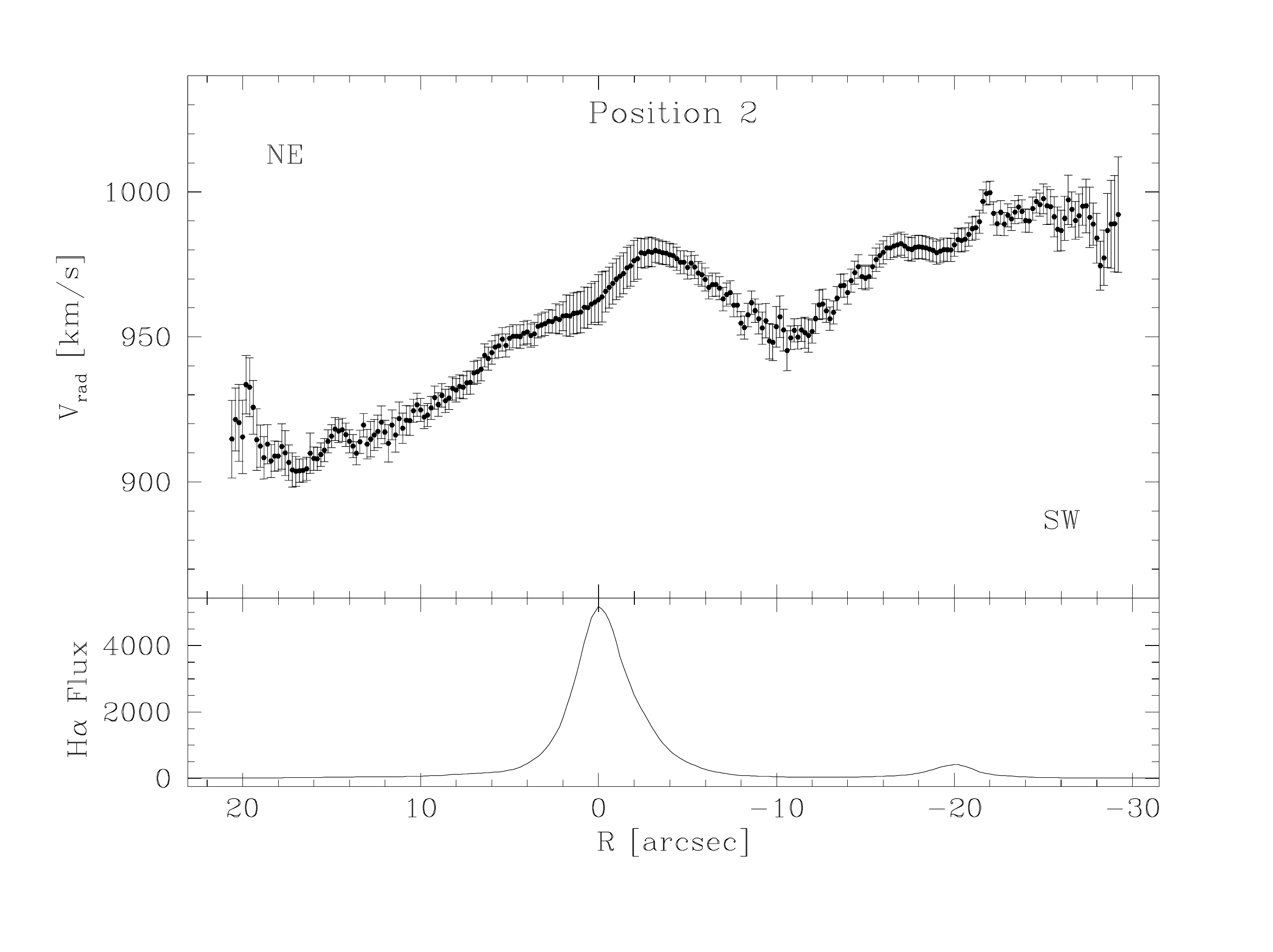}
\caption{Velocity profile for Position 2. Coordinate $R=0$ was set at the
center of knot \knotA.} 
\label{Fig:kin_pos2}
\end{figure*}

The velocity profiles in the two slit positions are very similar
(Figures~\ref{Fig:kin_pos1} and \ref{Fig:kin_pos2}). There seems to be an
overall, regular rotation in the outer parts of the starburst region, with
amplitude of about 100 km s$^{-1}$, on which a seemingly counter-rotating
component, with  a size of about $10\arcsec$, is superposed. This decoupled
region is located southwest of the continuum  and \Ha\ peaks, and can be
identified with the depression in gas emission between the central regions and
the two southwest knots, \knotC\ and \knotD\ (see Fig.~\ref{Fig:slitpos}).

\subsubsection{Emission Line Contribution to the Broad-Band filters}
\label{sect:EmLinContr}

In objects with significant gas emission, such as BCDs, emission lines can
contribute significantly to the flux through a broad-band filter, and thus
affect broad-band colors; the exact amount of such contribution in a given 
band depends on the intensity of the emission lines and on their location 
under the filter transmission curve. Most of the evolutionary synthesis models
(e.g. {\sc starburst99}~) do not take into account the emission line
contribution to the broad-band filters, and therefore it is essential that
this contribution be measured and subtracted out before computing the
broad-band colors of the knots.

To do this we proceeded as follows. At each spatial position, and for each
filter, we computed the integral of the curve obtained by multiplying the
number of photons emitted per unit wavelength interval in the observed 
spectrum by the filter transmission curve. 
The same calculation was repeated on the continuum, modeled by fitting a 
polynomial to the spectrum (we found that orders of about 20 gave the
best results), using an iterative sigma-clipping algorithm that rejects all 
datapoints more than about 1 sigma off the fit.

This was straightforward for the B and V filters, as their cutoff values fall
inside the wavelength range of our spectra. As for R, whose transmission curve 
extends to longer wavelengths than our spectra (which terminate at 
$\lambda \simeq 6920$), we computed $R'$ magnitudes, which we define
as the magnitude for a $R'$ filter whose transmission curve is equal to 
that of the standard R filter up to $\lambda \simeq 6920$, and zero 
afterward.

By computing the ratio between the two integral values and transforming into
magnitudes, we could compute the magnitude differences $B_{\rm obs}-B_{\rm
cont}$, $V_{\rm obs}-V_{\rm cont}$ and $R'_{\rm obs}-R'_{\rm cont}$, as well
as the offsets in color $(B-V)_{\rm obs}-(B-V)_{\rm cont}$ and $(V-R')_{\rm
obs}-(V-R')_{\rm cont}$. Note that such values are independent of any
calibration constant.

\begin{figure*}   % Figure 
\includegraphics[angle=0,width=\textwidth]{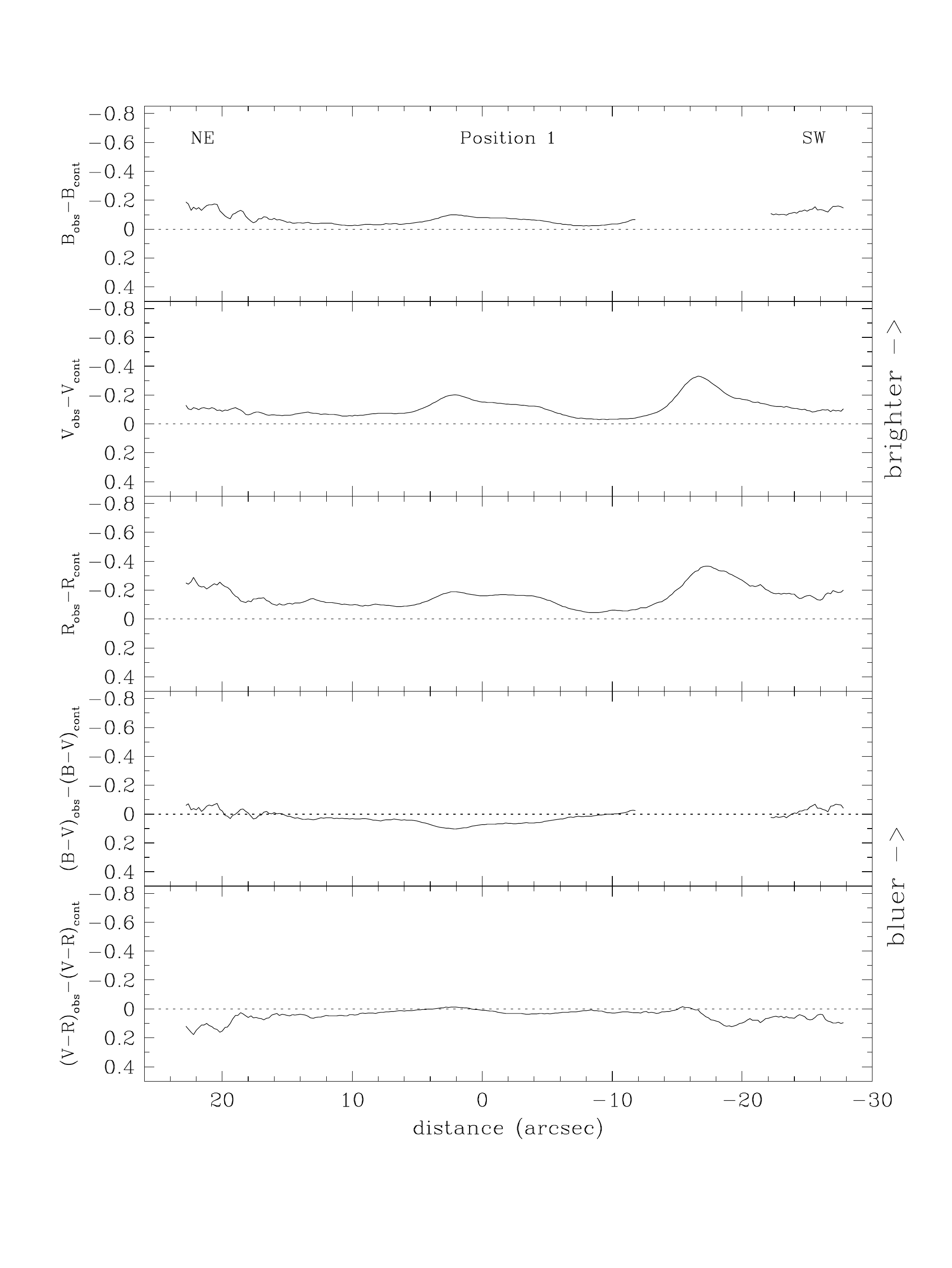}
\caption{Magnitudes and colors differences between the observed spectrum
and the interpolated continuum for Position~1. The discontinuities in the
$B$ and \BV\ plots correspond to the region affected by the 
''[\ion{O}{2}] ghost'', as described in \S~\ref{Section:PhysicalConditions}.} 
\label{Fig:magpos1}
\end{figure*}
\begin{figure*}   % Figure 
\includegraphics[angle=0,width=\textwidth]{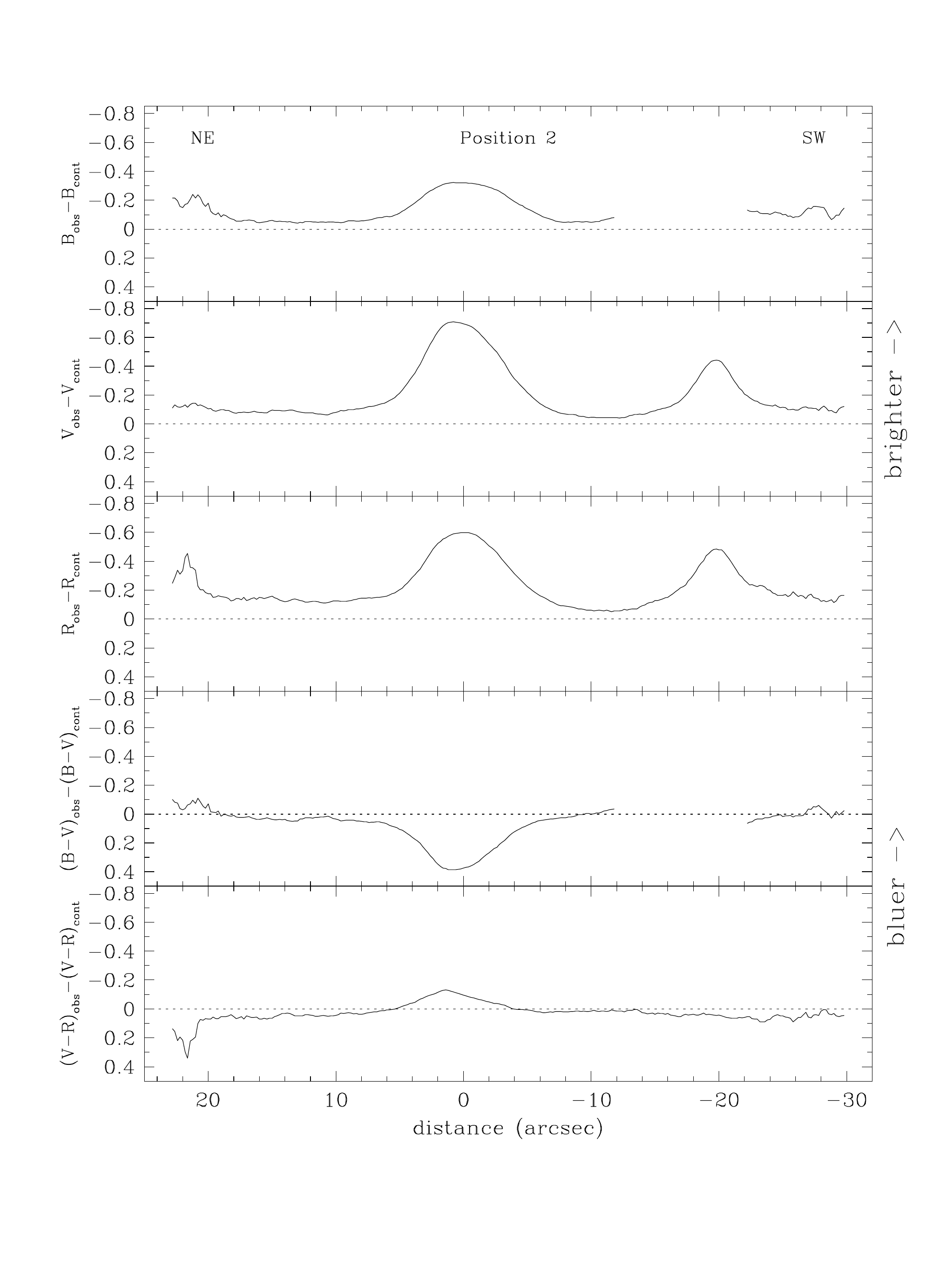}
\caption{Magnitudes and colors differences between the observed spectrum
and the interpolated continuum for Position~2.} 
\label{Fig:magpos2}
\end{figure*}

Figs.~\ref{Fig:magpos1} and \ref{Fig:magpos2} display the magnitude difference
as well as the color offsets, along the two slit positions. From these
figures it is evident the importance of performing a spatial correction: by
ignoring the correction or by adopting some averaged  values, we can obtain
misleading results.  For instance, along position 1 (see 
Fig.~\ref{Fig:magpos1}), the emission line contribution is $\lesssim 0.2$ mag
in $V$, except for knot \knotC\ (at $r \simeq 18$ arcsec), where it increases 
to $\simeq 0.35$. Along position 2, the emission line contribution is larger 
(with a peak of $\simeq 0.70$ in $V$ in knot \knotA) and shows greater 
variations.

Unfortunately as we do not have NIR spectra, we could not extend this analysis
to the NIR domain.

\subsection{Photometric analysis}

\subsubsection{Morphological view of Mrk~35}

The B-band contour map and the \Ha\ map of Mrk~35 are displayed in
Fig.~\ref{Fig:slitpos}. The starburst knots are aligned along the NE--SW
direction, forming a bar-like structure, while the brightest knots 
(\knotA, \knotBone\ and \knotBtwo ) form a "heart-shaped" structure in the
central part of the galaxy. \knotBtwo\ is the center of the outer isophotes,
while knot \knotA\ is the brightest SF region in the galaxy; the peaks of 
\knotA\  and \knotBtwo\ are separated by a distance of $3\farcs9$ 
(about 300 pc); in  HST frames \citep{Johnson04} both knots are resolved into 
a number of smaller  SF regions, many of them probably Super Star Clusters 
(SSCs). 
The optical  frames suggest the presence of a dust lane (crossing the galaxy
between knot \knotA\ and knot \knotB, and extending out to the west), which is
more clearly  visible in the high resolution HST/WFPC F606W image by
\cite{Malkanetal98}. The dust lane is more prominent in the $U$ and $B$
frames, and fade away  moving toward redder wavelengths; it displays a redder
color than the LSB component.

A tail-like feature extends in the SW direction, and connects with knots
\knotC\ and \knotD, located $\approx20\arcsec$ (1.5 kpc) from the center.  
The total length of the SF bar-like structure is $\approx 70\arcsec$ (5.3
kpc); the equivalent radius of the isophote at 27 $B$-mag arcsec$^{-2}$ is
$56\arcsec$ (4.2 kpc).

The isophotes twist going from the inner regions, dominated by the starburst,
to the outskirts of the galaxy, where they trace the shape of the underlying 
host, and have constant ellipticity for radii $\geq 30\arcsec$.  
The major axis of the outer isophotes has a position angle of  $90$ 
degrees, and forms an angle of $42$ degrees with the central chain 
of knots. 
\cite{Sanchez00} reported a P.A. of $\approx 45$ degrees for the outer 
isophotes, while \cite{Steel96} set the major axis of the outer isophotes at 
P.A. $\approx 58$ degrees. In both cases the photometry was not deep enough 
to reach the low surface brightness (LSB) galaxy host, and what is claimed 
to be the ``outer parts'' is actually a region where the starburst emission 
is still significant. This discrepancy stresses once more the need for deep, 
high-quality data to correctly characterize the host galaxy in BCDs.

The morphology of the galaxy in the broad-band optical frames basically
coincides with that in the infrared, though in the optical the inner isophotes
are more distorted, likely due to dust. The spatial position of the five
stronger sources coincide in all the broad-band frames. Knots \knotBone\
and \knotBtwo\ are the intensity peaks in the NIR, but they are not strong
emitter in the blue band; and while \knotA\ is the intensity  peak in the
bluer bands ($U$, $B$, $V$), it is a weak source in the NIR.  
This fact indicates a highly inhomogeneous dust distribution and/or age 
differences among the knots.

Fig.~\ref{Fig:slitpos}c is the gray-scale \Ha\ map of Mrk~35; $B$-band 
contours are over-plotted. The \Ha\ emission is located in the central regions
of the galaxy, elongated in the NE-SW direction, but slightly off-center to
the NW. The underlying stellar host extends out much farther than the
starburst region. The peaks of knots \knotA, \knotC\ and  \knotD\ in the
continuum (broad-band  frames) are also peaks in \Ha, whereas knot \knotBtwo, 
which is a strong emitter in the continuum, do not correspond to any of the 
\Ha\ local maxima. The \Ha\ map reveals intriguing features. Besides the eight
relatively large SF regions labeled in Fig.~\ref{Fig:slitpos}, we found a
number of smaller condensations around the main body of the galaxy: three
knots, together with knot \knotE\ form a ''chain-like'' structure extending in
a north-east direction; two other  knots are located about $10\arcsec$ north
of knots \knotC\ and \knotD; two small condensations are found south of knot
\knotD\footnote{These knots are faint and not visible in the \Ha\ maps
presented in Fig.~\ref{Fig:slitpos}}. All these knots are continuum emitters,
so they are stellar clusters. 
Faint, diffuse extensions emerge in the form of plumes and arcs from the main 
body of the galaxy; in particular, a remarkable arc-like structure with a 
diameter about 1 kpc, departs (apparently from knot \knotA) towards the West.

\subsubsection{Color maps}

High resolution and deep multi-wavelength broad-band observations, from the  
blue to the NIR, complemented by narrow band imaging, are powerful tools 
to study the stellar content of such complex systems as BCDs: 
i) the narrow-band frames allow us to identify the regions of the galaxies 
where active star-formation is taking place and, at the same time, 
the areas free of gas emission;  
ii) in the color maps we can discriminate the different stellar populations 
and see what regions are affected by gas contamination (for instance in \BV\
or \BR\ maps) or by dust lanes and patches (optical--NIR color maps).

\begin{figure}[h]   
\begin{center}
\includegraphics[angle=90,width=0.5\textwidth]{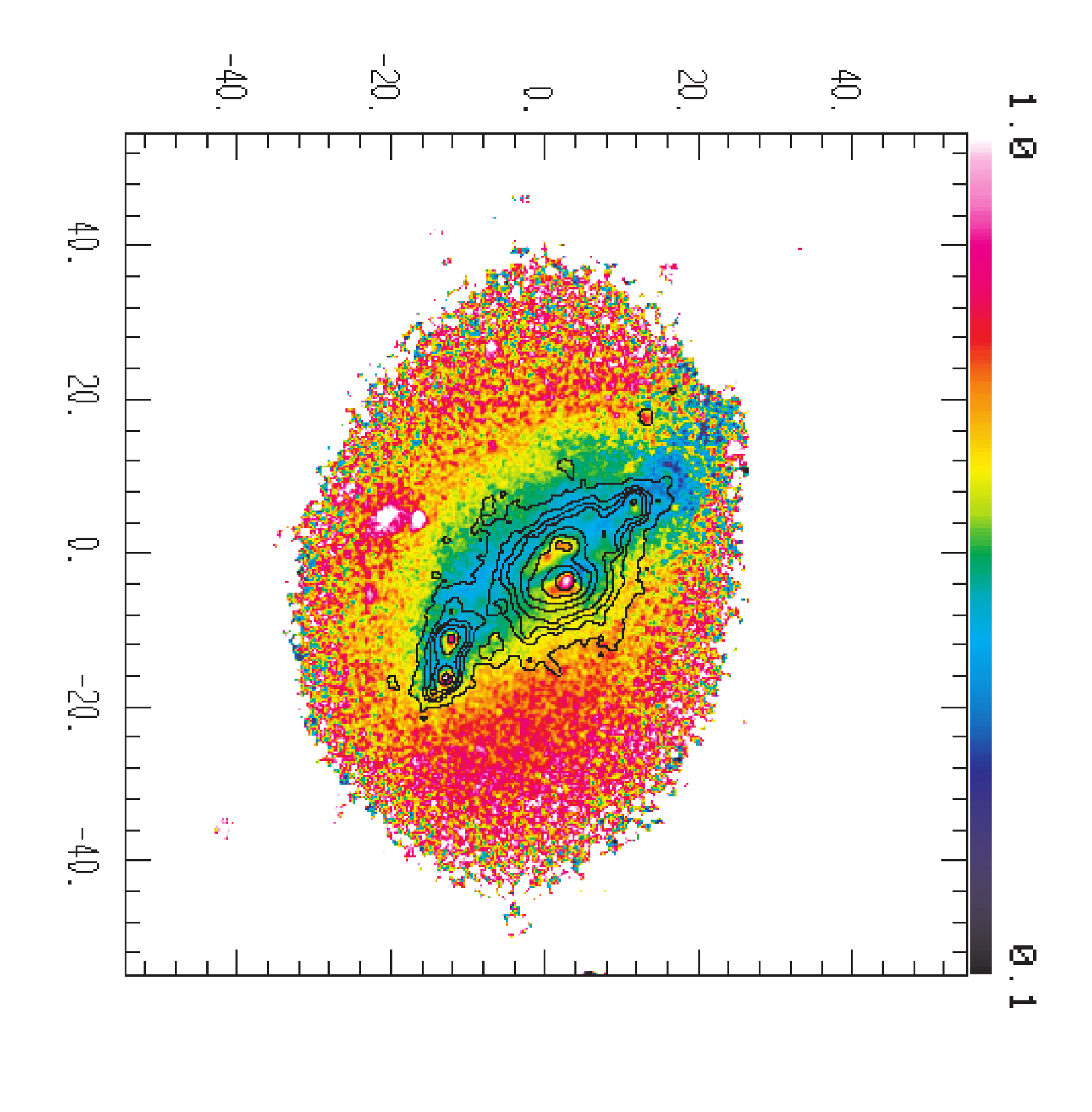}
\caption{$B-V$ color map of Mrk~35; isocontours for the continuum subtracted 
\Ha\ frame are overlaid. North is up and east to the left; axis units are 
arcseconds.}
\label{Fig:bvcolormap}
\end{center}
\end{figure}

Fig.~\ref{Fig:bvcolormap} displays the \BV\ map of Mrk~35; we clearly
distinguish the starburst, elongated in the north-east south-west direction and
placed atop the underlying extended and red elliptical host. The color
distribution of the starburst region is neither blue nor homogeneous, as red
patches appears along the chain. This is due to the emission line
contamination: the strong [\ion{O}{3}]~$\lambda\lambda$4959, 5007 lines account for a
large fraction of the flux through the V filter, whereas  the contribution of
the emission lines in the $B$ filter is significantly  smaller, producing a
substantial reddening of the $B-V$ color (see Figs.~\ref{Fig:magpos1} and
\ref{Fig:magpos2}).  We have over-plotted the \Ha\ emission on the color map;
notice that the  reddest patches inside the starburst coincide with the 
stronger gas emitters  (namely, \knotA, \knotC\ and \knotD). This result 
highlights, once again, the  need of taking into account the contribution of 
the gas emission and internal extinction, when interpreting the broad-band 
colors in starbursts.

\begin{figure}[h]   
\begin{center}
\includegraphics[angle=90,width=0.5\textwidth]{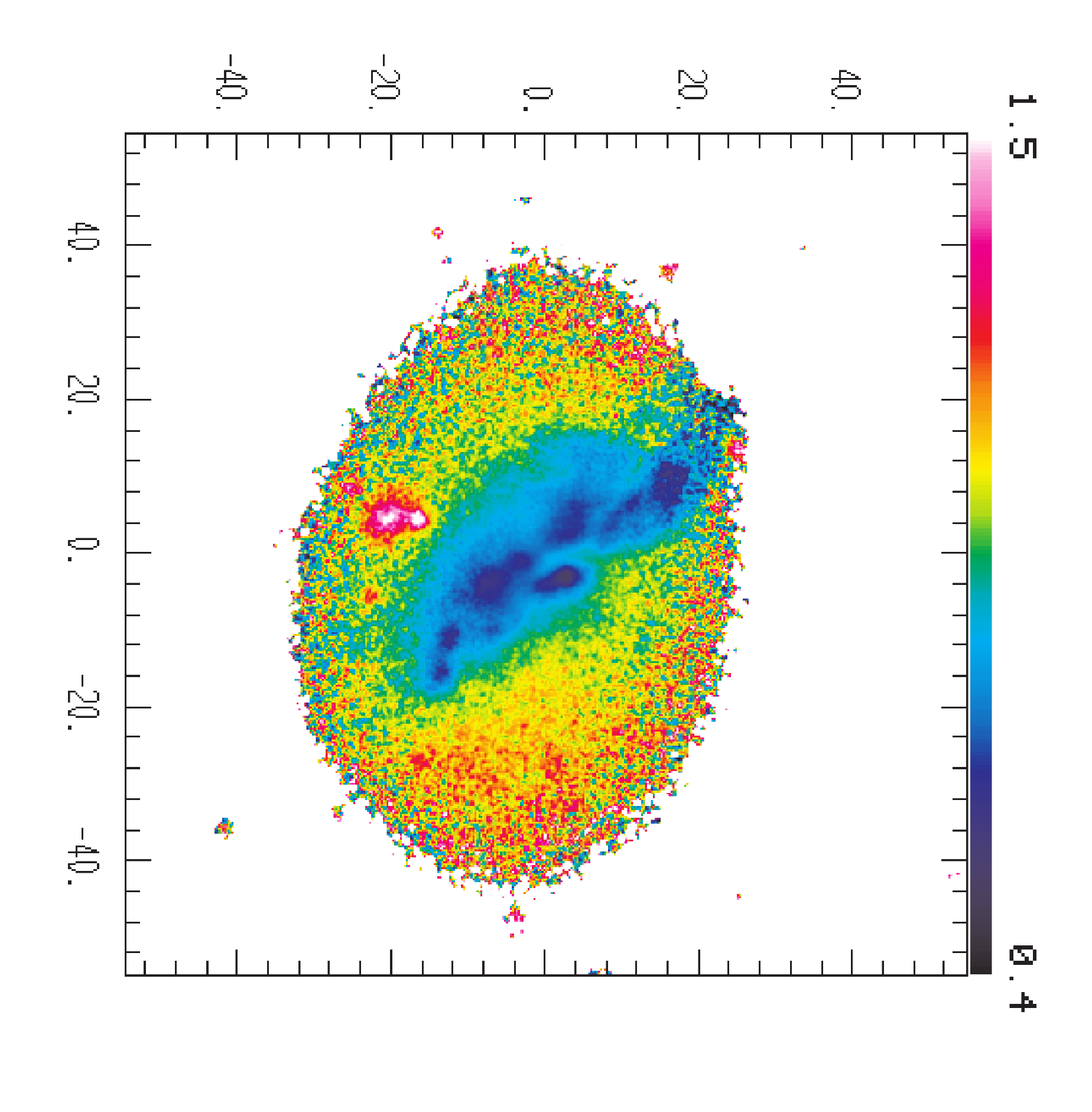}
\caption{\BR\ color map of Mrk~35. North is up and east to 
the left; axis units are arcseconds.}
\label{Fig:brcolormap}
\end{center}
\end{figure}

\begin{figure}[h]  
\begin{center}
\includegraphics[angle=90,width=0.5\textwidth]{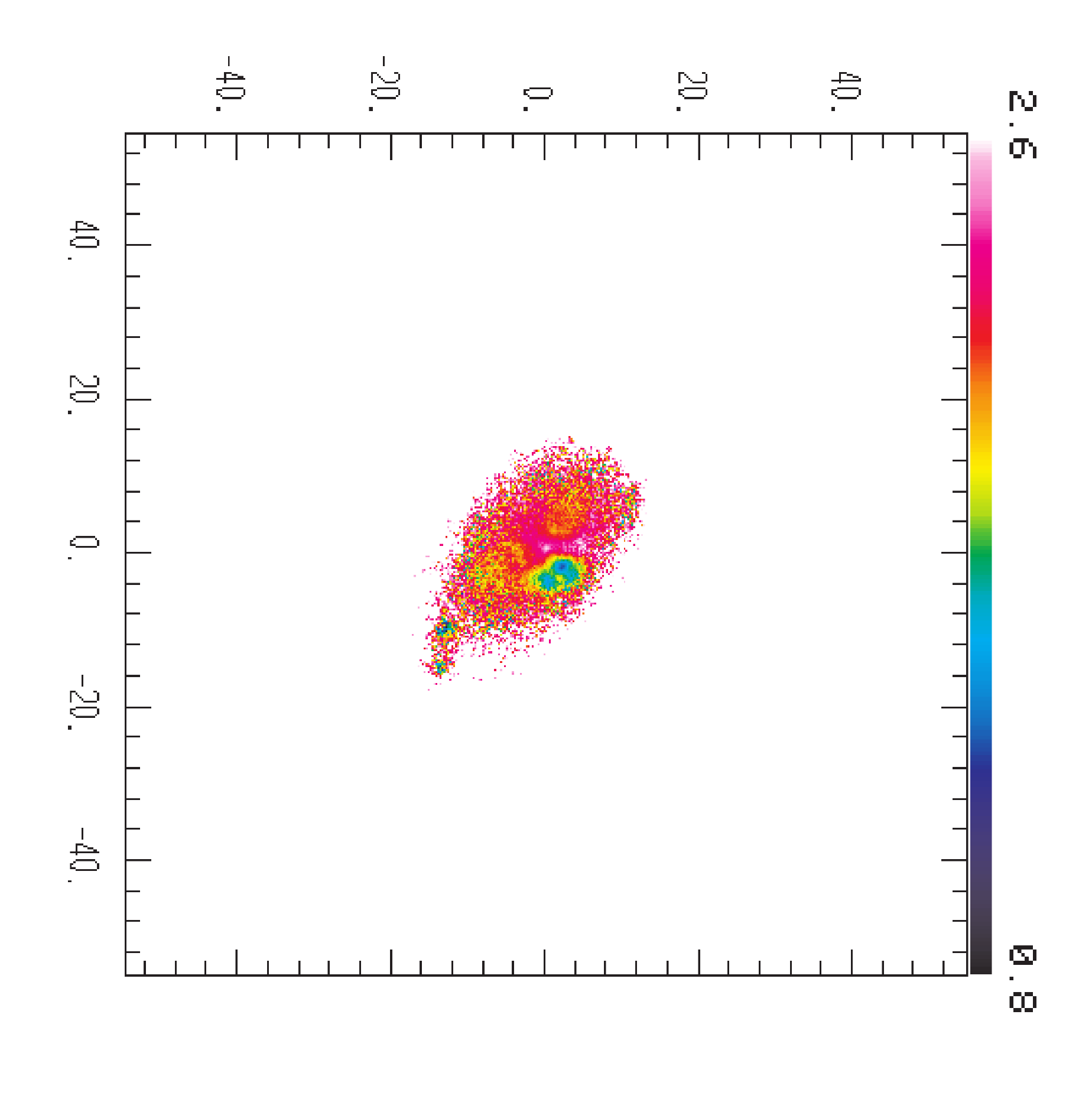}
\caption{\VK\ color map of Mrk~35. North is up and east to the left; 
axis units are arcseconds.}
\label{Fig:vkcolormap}
\end{center}
\end{figure}

Figs.~\ref{Fig:brcolormap} and \ref{Fig:vkcolormap} display the \BR\ and 
\VK\ color maps of Mrk~35. Because of the large difference in the interstellar
extinction coefficient between blue and infrared bands, these maps are useful 
to recognize and trace dust lanes and patches. Unfortunately, the $K$ frame  
is not as deep as the optical, and so we can build reliable color maps only 
for the central regions of the galaxy.  A red band, crossing the galaxy in the
north-south direction, between knots \knotA\ and \knotB, is visible: this is 
possibly a dust lane.
%
% Esto me convence poco, primero no veo ninguna banda roja, segundo porque es
% probable que sea una dust lane?
%
The inner regions in these maps are unlike the broad-band or the \Ha\ knot
distribution, as they display a patchy pattern, which probably results from 
the mixture of gas and dust.

\subsubsection{The Low Surface Brightness Stellar Component}
\label{Sect:hostmodelling}

Surface photometry of Mrk~35 in the optical and in the NIR was presented in
C01b, C03 and \cite{Caon05}. In these works, the 1-D luminosity profile of the
LSB component was modeled by a \Sersic\ law \citep{Sersic68}.  
Here, we carry out a two-dimensional modeling of the surface 
brightness distribution of the LSB component following the method described 
in \cite{Amorin07}. The two main advantages of this novel approach are: the 
possibility of masking out the starburst region by following exactly its 
shape, rather than setting an inner radial limit to the 1-D profile fit; and 
the fact that we can then subtract out the best fit model from the galaxy 
image to recover the starburst. 

The fit is done on the outer parts of the galaxy after masking out the 
central starburst and other disturbances such as small knots, foreground and 
background objects, etc. We used both H$\alpha$ images and color maps to
measure the size of the central starburst region. We start with a first mask,
and run the program. Discrepant points are flagged out and the fit is
re-done. The process is iterated until the fit converges, within a preset
tolerance, to a solution that does not depend on the exact values of the
input parameters.

The resulting two-dimensional \Sersic\ parameters of Mrk~35 in the optical and
in the NIR bands are shown in Table~\ref{Table:ajuste}. The model built in
this way is finally subtracted from the original galaxy frame to recover the
young starburst.

\begin{deluxetable}{lcccccccccc}
\tabletypesize{\footnotesize}
\tablewidth{0pt}
\tablecaption{2D S\'ersic photometric parameters for the LSB host galaxy
\label{Tab:ajuste}}
\tablehead{
\colhead{Band} & \colhead{$R_{\rm mask}$} & \colhead{$n$}  &
        \colhead{$R_{\rm e}$} & \colhead{$\mu_{\rm e}$} &
        \colhead{$m_{\rm LSB}$} & \colhead{$b/a$} &
        \colhead{P.A.} & \colhead{$c$} \\
	(1) & (2) & (3) &
        (4)& (5) &
        (6) & (7) &
        (8) & (9)}
\startdata
B & 20.80 & 0.98 & 15.04 & 22.33 & 14.11 & 0.72 & 74.0 &    $-0.02$ \\
V & 19.45 & 1.01 & 14.72 & 21.58 & 13.41 & 0.71 & 77.3 & \phs$0.03$ \\
R & 19.45 & 0.97 & 14.94 & 21.28 & 13.11 & 0.70 & 77.7 &    $-0.01$ \\
I & 14.50 & 0.83 & 15.33 & 20.80 & 12.63 & 0.71 & 74.9 & \phs$0.01$ \\
J & 12.80 & 1.12 & 16.70 & 20.52 & 11.65 & 0.70 & 58.1 & \phs$0.18$ \\
H & 14.02 & 1.01 & 19.99 & 20.00 & 11.19 & 0.67 & 63.2 & \phs$0.15$ \\
K & 12.80 & 0.88 & 20.50 & 19.70 & 10.89 & 0.70 & 60.6 & \phs$0.03$ \\
\enddata
\tablecomments{
Column (1): filter; 
col.\ (2): equivalent radius of the mask covering the central starburst 
region (arcsec);  
col.\ (3): \Sersic\ shape index $n$ ; 
cols (4) and (5): effective radius (arcsec) and surface brightness (mag
arcsec$^{-2}$);  
col.\ (6): total magnitude of  the LSB component derived from the fit (mag); 
col.\ (7): axis ratio; 
col.\ (8): position angle; 
col.\ (9): diskiness (negative)/boxiness (positive) parameter.}
\label{Table:ajuste}
\end{deluxetable}

\subsubsection{Disentangling the light from the different components}
\label{Sect:disentangling}

The flux received from any given point in the galaxy is the sum of three
components: the low surface brightness host, the continuum flux from the young
stellar component and the flux in emission lines:
\begin{equation}
F_{\rm tot} = F_{\rm LSB} + F_{\rm young} + F_{\rm emlines}
\end{equation}

The photometric analysis allows us to compute $F_{\rm LSB}$ by means of 
the \Sersic\ modeling, as explained in the previous section, and thus derive
\begin{equation}
F_{\rm young} + F_{\rm emlines} = F_{\rm tot} - F_{\rm LSB}
\end{equation}

The spectroscopic analysis allows us to compute $F_{\rm LSB} + 
F_{\rm young}$ by modeling the spectral continuum as described in 
Sect~\ref{sect:EmLinContr}, and thus derive
\begin{equation}
F_{\rm emlines} = F_{\rm tot} - \left(F_{\rm LSB} + F_{\rm young}\right)
\end{equation}

Therefore, by combining photometry and spectroscopy we are able to derive
separately each of these three components in filters $B$, $V$, and $R$ (the 
latter if we agree that the difference between the $R'$ and the $R$
band is not significant for our purposes.)

We illustrate how we proceed by taking knot \knotA, whose diameter is about 
$6\arcsec$.
In \knotA\ the underlying stellar component, as described by the \Sersic\ 
model, accounts for $\simeq10$\% of the total flux in B and in V, and 5\% in 
R.
In the spectral aperture $6\arcsec$ wide centered on this knot, the 
continuum (that is, $F_{\rm LSB} + F_{\rm young}$) accounts for 75\% of the 
flux in $B$, the emission lines 25\%. In $V$ we have 55\% and 45\% 
respectively; in $R'$ 60\% and 40\%.

Combining the above data, under the simplifying assumption that the spectral
data are constant within the whole \knotA\ area, it is straightforward to
derive that the young stellar component, $F_{\rm young}$, accounts for 65\%
of the total flux in $B$.

The same procedure can be applied to the other knots. The results are
summarized in Table~\ref{Table:fluxratio}.

\begin{deluxetable}{lccc}
\tablewidth{0pt}
\tablecaption{Flux decomposition for Mrk~35 SF knots}
\tablehead{
  \colhead{Component}  & \colhead {B} & \colhead {V} & \colhead{R} }
\startdata
\multicolumn{4}{c}{Knot \knotA}                    \\
$F_{\rm emlines}$ &    24.4 &     44.9 &     39.8  \\
$F_{\rm LSB}$     &    10.5 &     10.0 &  \phn5.4  \\
$F_{\rm young}$   &    65.1 &     45.1 &     54.8  \\[3pt]
\multicolumn{4}{c}{Knot \knotB1}           \\
$F_{\rm emlines}$ &  \phn8.0 &    16.0 &     15.2  \\
$F_{\rm LSB}$     &  \phn8.5 & \phn8.2 &  \phn4.1  \\
$F_{\rm young}$   &     83.5 &    75.8 &     80.7  \\[3pt]
\multicolumn{4}{c}{Knot \knotC}                    \\
$F_{\rm emlines}$ &     15.2 &    22.9 &     24.8  \\
$F_{\rm LSB}$     &     39.3 &    38.2 &     18.4  \\
$F_{\rm young}$   &     45.5 &    38.9 &     56.8  \\[3pt]
\multicolumn{4}{c}{Knot \knotD}                    \\
$F_{\rm emlines}$ &     19.2 &    29.7 &     32.6  \\
$F_{\rm LSB}$     &     36.9 &    34.0 &     16.4  \\
$F_{\rm young}$   &     43.9 &    36.3 &     51.0  \\[3pt]
\multicolumn{4}{c}{Knot \knotB2}                   \\
$F_{\rm emlines}$ &  \phn7.0 &    12.9 &     14.0  \\
$F_{\rm LSB}$     &  \phn3.6 & \phn3.3 &  \phn3.9  \\
$F_{\rm young}$   &     89.4 &    83.8 &     82.1  \\[3pt]
\enddata
\tablecomments{Values are normalized to a total flux $F_{\rm tot}=100$.}
\label{Table:fluxratio}
\end{deluxetable}

\subsubsection{Photometry of the starburst}

In order to identify the individual SF knots we used the \textsc{focas} package
and applied it to the continuum-subtracted \Ha\ images. \textsc{focas} looks 
for local maxima and minima in the counts of the pixels in the image; we 
adopted the criteria that, for a knot to be detected, its area must  be 
larger than 38 pixels (corresponding to an equivalent diameter of  
$\approx 1\farcs3$ or 100 pc), to ensure that the diameter of the knot is 
larger than the point spread function; also the counts of each of its pixels 
must be higher than 2.5 times the standard deviation of the sky background. 
Using these criteria eight knots were identified (see 
Fig.~\ref{Fig:slitpos}b). 

The total \Ha\ flux of the galaxy is $1.09 \times 10^{-12}$ \ergscms\  (after
setting a threshold of $3.8\times 10^{-18}$ \ergscms, 2.5 times the standard
deviation  of the sky background). With a diameter of 0.64 kpc and a total 
flux of $5.57 \times 10^{-13}$ \ergscms, \knotA\ is the brightest \Ha\ knot,
emitting half of the total \Ha\ flux of the galaxy.

Next, we computed the broad-band magnitudes of the starburst knots identified 
in the \Ha\ frame, after convolving the broad-band images to match the PSF of
the image with the worst seeing. Results of the photometry are presented in
Tables~\ref{Table:hapho} and ~\ref{Table:optbb}. 
Table~\ref{Table:hapho} displays the \Ha\ flux, luminosity and equivalent width
of the knots. The central knot \knotBtwo, not detected by \textsc{focas}, is
also included (fluxes are computed within a circular aperture $1\farcs5$ in
radius). The data have been corrected from interstellar extinction and 
[\ion{N}{2}] contribution using our spectroscopic information. 

Table~\ref{Table:optbb} lists the broad-band colors of the knots corrected for
Galactic extinction, together with the "true" starburst colors, computed by
subtracting the contribution of the underlying host galaxy (see
Sect~\ref{Sect:hostmodelling}) and of the flux due to emission lines (for the
$B$, $V$ and $R$ filters, see Sect~\ref{Sect:disentangling}), and by correcting
for interstellar reddening. For those knots whose position matches a
spectroscopic  aperture (namely \knotA, \knotBone, \knotBtwo, \knotC\ and
\knotD) we correct from reddening and emission lines using the values derived
for the corresponding aperture; in the remaining cases we use the values 
derived for the nearest subregion in the spectrum.

\begin{deluxetable}{rcccc}
\tablewidth{0pt}
\tablecaption{\Ha\ photometry of the individual knots in Mrk~35}
\tablehead{
\colhead{Knot} & \colhead{Flux(\Ha)} &
            \colhead{$\log L(\Ha)$} & 
            \colhead{$-W(\Ha)$} &
            \colhead{$-W(\Ha)_{\rm cor}$} }
\startdata
A   &        556.60   & 40.21   &    450  &    572  \\
B1  &     \phn44.54   & 39.11   &    157  &    178  \\
C   &     \phn23.98   & 38.84   &    309  &    687  \\
D   &     \phn15.86   & 38.66   &    361  &    714  \\
E   &  \phn\phn0.27   & 36.90   &    151  & \nodata \\
F   &  \phn\phn6.60   & 38.28   &    166  &    244  \\
G   &  \phn\phn2.02   & 37.77   & \phn72  & \phn95  \\
H   &  \phn\phn6.04   & 38.24   &    102  &    143  \\
B2  &  \nodata  & \nodata & 124  & 140 \\                   
\enddata
\tablecomments{Values are corrected from interstellar extinction and from
[\ion{N}{2}] emission.
The equivalent widths are shown before and after the correction from the
contribution of the LSB component to the continuum.
\Ha\ fluxes are in $10^{-15}$ \ergscms\ units;
\Ha\ luminosities in  erg sec$^{-1}$ units. $W(\Ha)$ is in \AA.}
\label{Table:hapho}
\end{deluxetable}

\begin{deluxetable*}{lccccccccc}
\tabletypesize{\footnotesize}
\tablewidth{0pt}
\tablecaption{Observed broad-band photometry of the individual knots detected 
in Mrk~35. \label{Tab:mkn35bb}}
\tablehead{
\colhead{Knot} &
\colhead{Area} & \colhead{$B$} & 
\colhead{\BV}  & \colhead{\VR} & 
\colhead{\VI}  & \colhead{\VJ} &
\colhead{\JH}  & \colhead{\HK}}
\startdata
\multicolumn{9}{c}{Total magnitudes and colors} \\[4pt]
%       Area     B-mag    B-V     V-R           V-I      V-J      J-H         H-K
A   &    25.7 & 15.43  & 0.66	& 0.01    &  0.15     &  0.92	&  0.41   &   0.49  \\
B1  & \phd4.6 & 17.21  & 0.64	& 0.12    &  0.53     &  1.31	&  0.42   &   0.46  \\
C   & \phd8.4 & 18.61  & 0.61	& 0.09    &  0.35     &  0.98	&  0.50   &   0.31  \\
D   & \phd5.4 & 19.30  & 0.67	& 0.07    &  0.30     &  1.00	&  0.49   &   0.43  \\
E   & \phd2.4 & 22.33  & 0.67	& 0.26    &  0.78     & \nodata & \nodata & \nodata \\
F   & \phd4.9 & 19.38  & 0.49	& 0.20    &  0.54     &  1.27	&  0.38   &   0.42  \\
G   & \phd2.7 & 19.09  & 0.50	& 0.23    &  0.62     &  1.33	&  0.47   &   0.35  \\
H   & \phd5.6 & 18.61  & 0.49	& 0.25    &  0.63     &  0.95	&  0.49   &   0.39  \\
B2  & \phd7.1 & 16.89  & 0.63	& 0.14    &  0.70     &  1.43	&  0.49   &   0.42  \\
\hline\\[-3pt]
\multicolumn{9}{c}{Magnitudes and colors for the young component} \\[4pt]
%       Area     B-mag    B-V     V-R           V-I      V-J      J-H         H-K
A    &        &  15.68 & 0.22   & 0.18    &   $-0.01$ &  0.66   & 0.45    &   0.55  \\
B1   &        &  17.27 & 0.50   & 0.17    & \phs0.46  &  1.15   & 0.46    &   0.49  \\
C    &        &  19.37 & 0.41   & 0.48    & \phs0.04  &  0.62   & 0.47    &   0.44  \\
D    &        &  20.10 & 0.43   & 0.41    &   $-0.02$ &  0.70   & 0.35    &   0.66  \\
E    &        &  24.40 & 0.75   & \nodata & \phs0.85  & \nodata & \nodata & \nodata \\
F    &        &  19.66 & 0.23   & 0.42    & \phs0.40  &  0.78   & 0.47    &   0.55  \\
G    &        &  18.25 & 0.35   & 0.36    & \phs0.54  &  1.02   & 0.60    &   0.43  \\
H    &        &  18.87 & 0.22   & 0.47    & \phs0.54  &  0.62   & 0.62    &   0.46  \\
B2   &        &  16.83 & 0.52   & 0.08    & \phs0.63  &  1.28   & 0.52    &   0.44  \\
\enddata
\tablecomments{Upper rows: broad-band photometry for the star-forming 
regions detected in Mrk~35. Aperture photometry (within a radius = $1\farcs5$)
for knot B2 is also included. 
Integrated total magnitudes have been corrected for Galactic extinction.
Bottom rows: Magnitudes and colors of the young component, obtained by  
subtracting the low surface brightness host in every filter, and the 
contribution of emission lines in $B$ and in \BV\ and \VR\ colors. 
Magnitudes and colors have also been corrected for both Galactic and 
interstellar extinction.}
\label{Table:optbb}
\end{deluxetable*}

\subsection{Disentangling the Stellar Populations}

\subsubsection{The Starburst}

In order to constrain the properties of the SF knots identified in Mrk~35, we
adopted the {\sc Starburst99} evolutionary synthesis models 
\citep{Leitherer99}, which produce a comprehensive set of 
observables of galaxies with active star formation. We use models with
$z=0.004$ and $z=0.008$, which bracket the metallicity derived from the 
emission line fluxes (which is a reasonable approximation to the metallicity of
a young population), and look for the scenario that best reproduces the 
collection of observables available for every knot: $W(\Ha)$, the optical
and NIR colors, as well as $W(\Hb)$ for those knots with spectroscopic
information.  

In all cases, we can reproduce the \Ha\ equivalent width with an instantaneous
burst of star formation and the Salpeter {\sc imf} with an upper mass limit of
100 $M_{\odot}$. Knot \knotA, \knotC\ and \knotD\ are the youngest; their
observables are consistent with ages of about 5 Myr. The central knots,
\knotBone\ and \knotBtwo, are slightly older, with ages of about 7 Myr.

\begin{figure}[h]   
\begin{center}
\includegraphics[angle=270,width=0.5\textwidth]{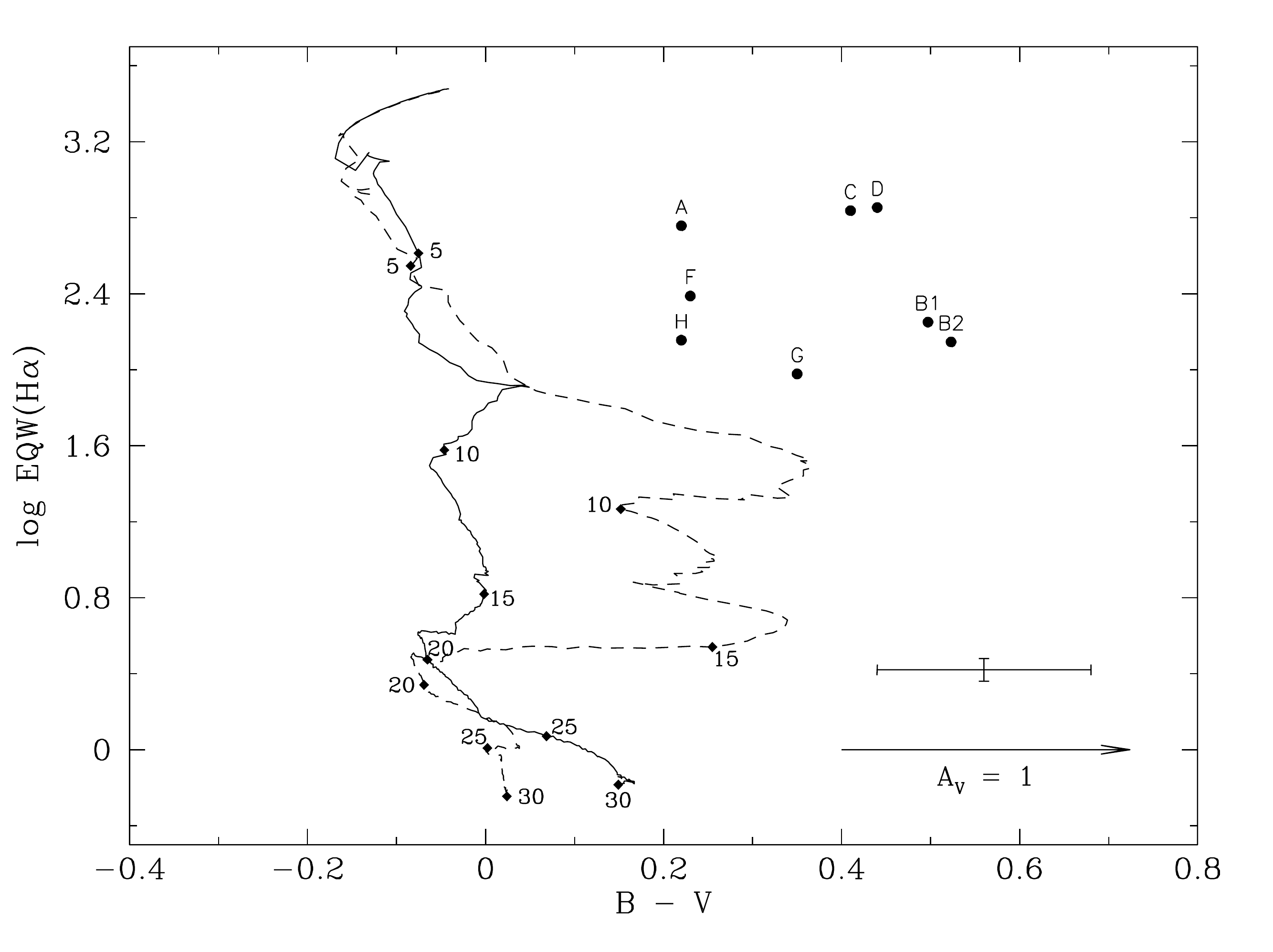}
\caption{Logarithm of the equivalent width of the \Ha\ line {\em versus} 
$(B-V)$.
Solid points are the observables of the the knots after applying the 
corrections from reddening and from the contribution of emission lines and 
older stars.
We also plot the tracks of two instantaneous bursts of $z=0.004$ (solid line)
and $z=0.008$ (dashed line), both aged from 1 to 30 Myr, and with an 
IMF with $\alpha=2.35$ and $M_{\rm up}=100M_{\odot}$ (Leitherer et al. 1999).
The numbers along the tracks mark the age in Myr.}
\label{Fig:modelos1}
\end{center}
\end{figure}

In Fig.~\ref{Fig:modelos1} our results are compared with the models. The
equivalent width of the \Ha\ line is plotted versus \BV.  All the knots
have \Ha\ equivalent widths in good agreement with model predictions, with ages
ranging from 4 to 7 Myr, but lie far from the models in the \BV\ axis. 

This is not surprising.
Although the data have been corrected for the contribution of emission lines 
(Section~\ref{sect:EmLinContr}), the contribution of the LSB host 
(Section~\ref{Sect:hostmodelling}) and for interstellar extinction,  
the various assumptions and simplifications we made to be able to work out 
such effects (for instance, that C(H$\beta$) and the emission line 
contribution do not vary spatially within a same knot) result in significant 
uncertainties on the final values. 
Also, in Mrk~35, the presence of a substantial amount of dust can also play 
an important role.

\begin{figure}[h]   
\begin{center}
\includegraphics[angle=270,width=0.5\textwidth]{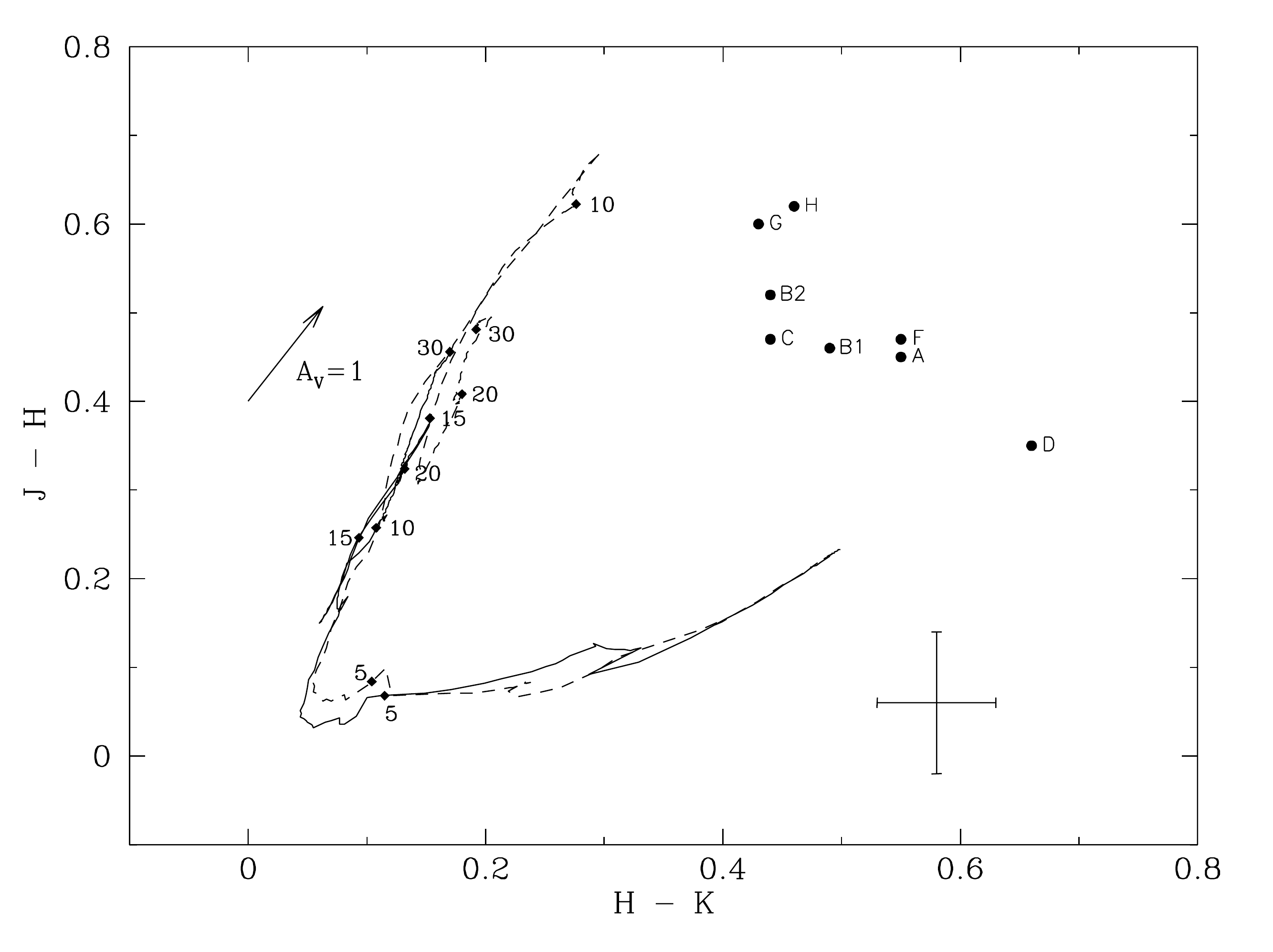}
\caption{\JH\ \emph{vs} \HK\ color-color diagram.
Solid points represent the knots parameters after applying the corrections 
from reddening and older stars.
We also plot the tracks of two instantaneous bursts of $z=0.004$ (solid line)
and $z=0.008$ (dashed line), both aged from 1 to 30 Myr, and with an 
IMF with $\alpha=2.35$ and $M_{\rm up}=100M_{\odot}$ (Leitherer et al. 1999).
The numbers along the tracks mark the age in Myr.}
\label{Fig:modelos2}
\end{center}
\end{figure}

Fig.~\ref{Fig:modelos2} displays the {\sc nir} color-color diagram of the SF
regions. All the observed knots lie off the region of the models. We must keep
in mind that {\sc nir} colors have not been corrected from gas line emission, 
and that SB99 includes only the contribution of the nebular continuum.  
However such a large disagreement is difficult to explain only in terms  of 
emission lines. 
These excesses in the IR colors, already observed in starburst  galaxies,
could be a signature of the presence of hot dust 
\citep{Vanzi02,VanziSauvage04,Johnson04,Cresci05}.

\subsubsection{The Low Surface Brightness Stellar Component}

To constrain the evolutionary status of the host galaxy in Mrk~35 we can only
use the information derived from its colors. Although some absorption features
are detected in the spectra, they cannot be measured with  sufficient accuracy.

The colors of the underlying galaxy host (derived from the \Sersic\ models) are
tabulated in Table~\ref{Table:hostcolors}. They are not corrected from internal
extinction, because the coefficients we derived, based on the gas emission
lines, may not apply to the regions outside the area occupied by the starburst.

\begin{deluxetable}{cccccc}
\tabletypesize{\footnotesize}
\tablewidth{0pt}
\tablecaption{Colors of the LSB stellar component derived by fitting a 
\Sersic\ model}
\tablehead{
\colhead{$B-V$} & \colhead{\VR} & \colhead{\VI}  &
        \colhead{\VJ} & \colhead{\JH} &
        \colhead{\HK}}
\startdata
0.70 & 0.30 & 0.78 & 1.76 & 0.46 & 0.30 \\
\enddata
%\tablecomments{}
\label{Table:hostcolors}
\end{deluxetable}

\begin{figure}[h]   
\begin{center}
\includegraphics[angle=0,width=0.5\textwidth]{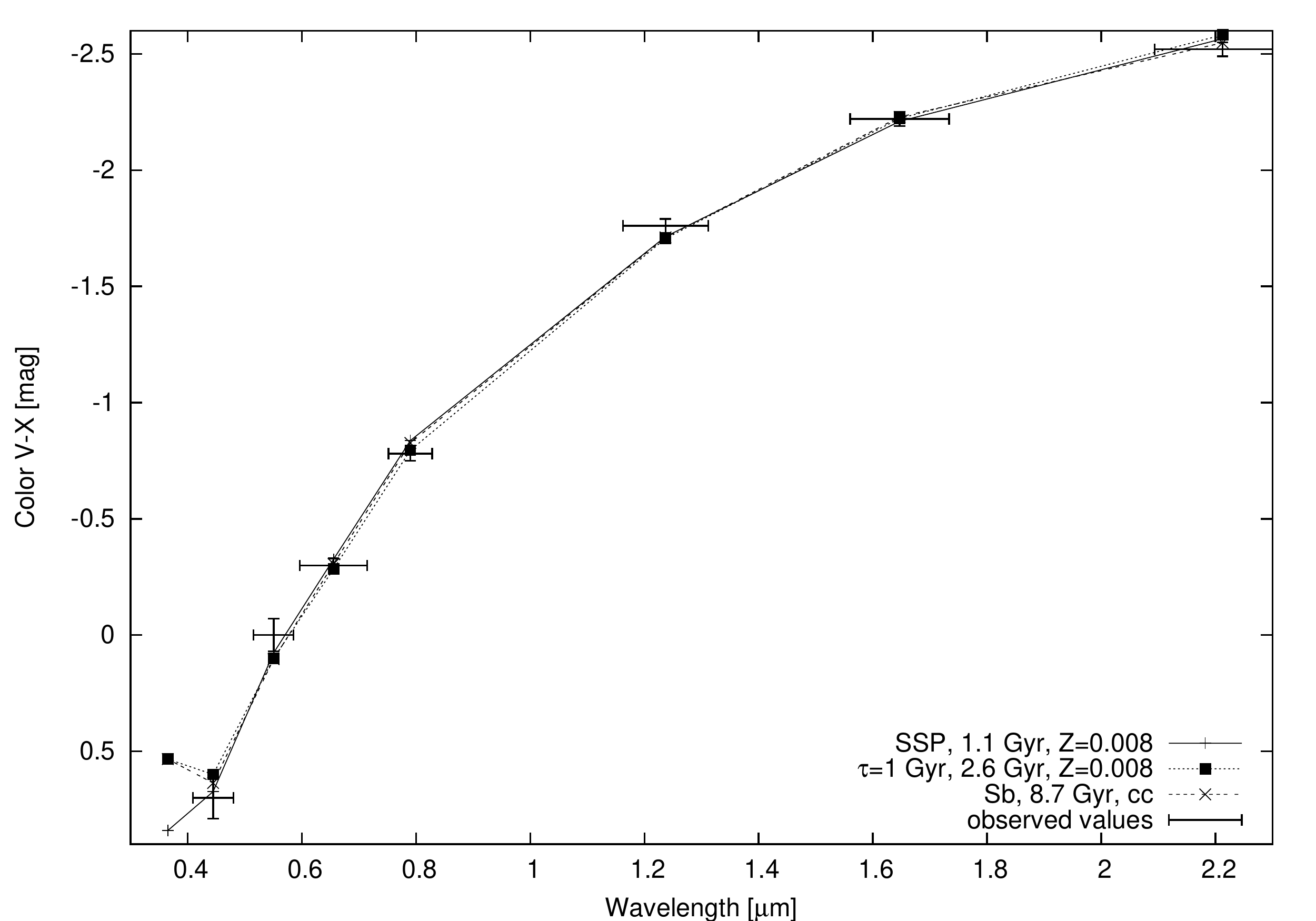}
\caption{The $B$ to $K$ spectral energy distribution of the observed LSB
component and three typical models. The y-axis shows the color relative to $V$,
the x-axis the wavelength. The observations are shown as crosses, where
the x-error gives the approximate width of the filters and the y-axis the
uncertainty of the S{\'e}rsic fit in each filter. Three selected models are 
shown: 1) an SSP with $Z=0.008$ at an age of 1.1\,Gyr; 2) a model with 
characteristic star-formation timescale $\tau=1$\,Gyr with $Z=0.008$ at an age 
of 2.6\,Gyr; 3) a chemically consistent model of an Sb galaxy 
($\tau=6.6$\,Gyr) at an age of 8.7\,Gyr.}
\label{fig:SEDcomp}
\end{center}
\end{figure}

We use the spectral energy distribution (SED) derived from the \Sersic\ model 
to estimate the age of the underlying old stellar component. We employ models
derived from the GALEV code \citep{Bicker04, Weilbacher04}; all of 
them use a Salpeter initial mass function and Padova isochrones and other 
input parameters as described by \cite{Anders04}. Our model grid consists of 
single metallicity models with $Z=0.004$ and $Z = 0.008$ and we use single 
stellar population (SSP) as well as models with exponentially decreasing
star formation rate with a characteristic timescale $1 < \tau < 12$\,Gyr. 
As an alternative we also include chemically consistent models similar to 
those of \cite{Bicker04}, which are divided into different star formation 
histories to represent all galaxy types observed at the present day and 
use exponential timescales from $\tau=1$\,Gyr to constant star formation rate.

To look for the best agreement between the measured colors and the models we
then derive an error weighted $\chi^2$ value of the observed magnitudes of
every time-step of each model. Several models give a good fit; all those with
half solar metallicity give a better fit than those with $Z=0.004$. It is,
however, difficult to obtain a good estimate of the dominant age of the
stellar populations. Despite the wide wavelength range covered, the possible
solutions are degenerate regarding age and star formation history (see
Fig.~\ref{fig:SEDcomp}). While the best matching {\it single} metallicity
model is the SSP with $Z=0.008$ at an age of 1.1\,Gyr, other models with
longer star formation timescales and only slightly worse $\chi^2$ values match
at ages of 2.5 to 7.9\,Gyr. Some of the chemically consistent models also give
very good fits, the Sb-galaxy modeled after \cite{Bicker04} provides the best
overall match at an age of 8.7\,Gyr.

We cannot help concluding that further observations, and especially data on 
spectral indices in the outer part of the LSB component, are necessary to 
determine the true dominant age of the underlying stars in the galaxy.

\section {DISCUSSION}

This work is the second of a series of papers devoted to a thorough
spectrophotometric analysis of a sample of nearby BCDs. As we concluded in our
previous work (C02, focused on the i,E BCD Mrk~370), this kind of analysis,
although scarce in the literature because of its difficulty, is essential to
disentangle the stellar populations in systems as complex as BCDs and to put 
strong constraints on their SF histories and evolutionary status.

Here, by combining a wide set of photometric and spectroscopic data we 
disentangled the two stellar populations in the iE BCD Mrk~35: the starburst, 
which forms a bar-like structure extending in the northeast-southwest  
direction, and the underlying population of substantially older stars.  
The current star-formation not only takes place  in the central knots 
(\knotA\ and \knotB), as pointed out by \cite{Johnson04} ---~who worked with 
frames that covered only the innermost regions~---, but also in the two 
knots placed at the end of the "tail-like" feature (\knotC\ and \knotD), as
their \Ha\ emission reveals. 

The brightest knot, \knotA, accounts for half of the \Ha\ emission of the 
galaxy; the age we derived for it is about 5 Myr, consistent with the
ages found in previous works \citep{Johnson04,Steel96}, and with the detection 
of WR stars \citep{Steel96, Huang99}. 
Knots \knotC\ and \knotD\ show almost identical ages, which rules
out the hypothesis that the star formation is propagating along the  
"tail-like" feature. 
Knots \knotBone\ and \knotBtwo\ have slightly larger ages, about 7 Myr, 
but the age difference with the brighter knots ($\sim2$ Myr) is probably too
small to be consistent with a scenario of self-propagating star-formation.

The high resolution frames combined with spatially resolved long-slit
spectroscopy allowed us to carry out an exhaustive analysis of the starburst
population.  In order to characterize the current starburst episode, a
considerable effort has been made to disentangle the young stars from the
ionized gas emission, from dust and from older stars. 
Using deep broad-band frames we modeled the LSB host and subtracted it from 
the total emission; the spectra allowed us to estimate the contribution of 
emission lines, and to derive the extinction coefficient, in the different 
spatial regions along the two slit positions (see 
Table~\ref{Table:fluxratio}). 

After applying these corrections, we derived colors, physical parameters and 
oxygen abundances for the SF knots, and investigated how they vary from one 
spatial location to another.

A very important aspect, often ignored in previous analyses, is the estimation
of the contribution of emission lines to the total amount of flux measured 
using standard broad-band filters. In this case, using our spectra we 
calculated the amount of flux due to emission lines in the the $B$, $V$ and $R$ 
bands, along both slit positions. We found that this contribution is by no
means negligible, and even more importantly, it shows spatial variations, not
only from one knot to another, but even within the same knot. For instance, the
contribution of emission lines in knot \knotA\ (along the slit position  2), is
as large as $0.7$ mag; such a huge amount of flux  can make the \BV\ starburst
colors similar or even redder than those of the host (see the \BV\ color map in
Fig.~\ref{Fig:bvcolormap}). Therefore, for a proper subtraction  of the emission
lines contribution a two-dimensional mapping of the line emission is  required.

Several recent works have questioned the previous belief that BCDs have no 
or little dust \citep{Hunt01,Cairos03,VanziSauvage04,Cabanac05}.
Mrk~35 shows evidence of dust in its central regions. In the optical frames 
dust is visible as a lane between the central knots, \knotA\ and 
\knotB, extending to the west (as already pointed out by C03 and 
\citealp{Johnson04}). 
Our color maps also suggest the existence of a more extended dust 
distribution: red patches are in fact visible at the north of knot \knotA, and 
the pronounced "red" border, crossing the galaxy parallel to the central 
like-bar starburst, may also be due to dust which blocks light from the 
part of the galaxy behind.

One intriguing aspect is that, although we found evidence for the presence 
of dust, the derived reddening values in all the apertures are considerably 
small.
From the Balmer decrement, we obtain $A_{V}=0.17$ in knot \knotA\ and
$A_{V}=0.11-0.13$ in knots \knotBone\ and \knotBtwo\ (in good agreement with 
the values reported in \citealp{Steel96}). \cite{Johnson04} however, working 
in NIR and radio wavelengths, found considerably higher extinction value for 
knot \knotA, $A_{V} = 8$. These discrepancies in the extinction measured at 
different wavelengths have been also found in other SF dwarfs (VV~114, 
\citealp{Yun94}; He2-10, \citealt{Kobulnicky95}): in a dusty environment, in 
which the dust is mixed with the emitting sources, there is a trend of 
increasing extinction from the optical to the IR. 
Therefore, the lower extinction values derived from our optical spectroscopy 
are most probably the result of a inhomogeneous dust distribution. 
% The excess that we found in the NIR colors is also consistent with the 
% presence of hot dust \citep{Lada92}.

Indeed, the actual amount of dust (and its distribution) is currently a hot
topic in BCD galaxy research. By studying the properties of the dust in these
nearby dwarfs we can learn about the dust formation in primordial 
environments. Moreover, it is becoming more and more clear that an 
important fraction of the SF activity in these galaxies lies buried in 
dust and has not visible counterparts \citep{Thuan99,KobulnickyJohnson99,
Hunt01,Vaccaetal02,Johnsonetal03,VanziSauvage04}. 
Integral Field spectroscopic observations, with which to build a
spatial map of the extinction coefficient, together with high-resolution 
MIR observations to map the dust distribution and measure its 
emission, would provide important advances in the field. 

The galaxy morphology ---~two major knots in the inner regions of  the galaxy
and the "tail-like" features~--- suggests that an  interaction (and/or) merger
could have played a main role in the galaxy  shaping and evolution.  
\cite{Steel96} rejected the merger hypothesis on the basis on the velocity  of
the central knots (almost identical) and the regularity of the outer isophotes,
and favored the hypothesis of intrinsic self-induced  star-formation.
Nevertheless, only small age gradients have been detected across the galaxy,
even with the deeper and better spatial resolution data in \cite{Johnson04} and
in this work. Although we cannot discard that some  propagation is operating in
Mrk~35, it does not seems to be the dominant mechanism.  On the other hand,
\cite{Johnson04} favored a merger event as the one igniting the star-formation
in Mrk~35: they argued that the regularity of the outer isophotes does not
exclude a small-scale interaction, and the  fact that \knotA\ and \knotB\ had
similar redshifts also does not exclude that they were once separated systems. 
They compared Mrk~35 with the iE, BCD He~2-10, a very well studied object
\citep{Kobulnicky95,Johnson00,Cabanac05}, which has significant  similarities
with Mrk~35, and appears to be interacting with a massive cloud of gas that is 
falling into the main body of the galaxy \citep{Kobulnicky95,Johnson00}, and
speculate that the nature of the star-forming trigger in Mrk~35 could be
similar. 

Our kinematical analysis seems to support the merger hypothesis. The apparently
counter-rotating feature in the inner region of the velocity profiles could be
the signature of a past merger or acquisition event. Integral
field spectroscopy, which allows the derivation of the velocity field of the 
gas, is required to properly interpret the rotation pattern. Also, 
deep spectroscopic observations with large telescopes, which permit high 
signal-to-noise observations of absorption features and therefore to trace 
the kinematics of the stars, would be essential.

\section{Summary and Conclusions}

Optical and NIR broad-band images, \Ha\ narrow-band frames  
and long-slit spectroscopic data for the i,E BCD galaxy Mrk~35 have
been analyzed in order to derive the properties of the different components of
the galaxy, to constrain its evolutionary status and to investigate the 
possible mechanisms triggering the actual SF episode. From this study we 
highlight the following results:  

\begin{itemize}

\item Two different stellar components are clearly distinguished in the galaxy:
the actual starburst and an underlying older population. The current star
formation activity takes place in several knots, aligned along the NE-SW
direction, in a bar-like structure; the brightest knots (\knotA, \knotBone\ 
and \knotBtwo\,) are arranged in a "heart-shaped" structure, in the central
region of the galaxy, while knot \knotC\ and \knotD\ are located in the
"tail-like" feature.  An extended LSB component, with regular appearance and 
red colors, sits underneath the star-forming area.

\item  The brightest knot, \knotA, is a powerful and young SF region, with
Wolf-Rayet stars, and accounts for 50\%  of the total \Ha\ luminosity of 
the galaxy.  

\item  For each spatial region in the slit we derived reddening corrected 
intensity ratios, extinction values, physical parameters of the gas and  
oxygen abundances. Although we found evidences for the presence of dust, the 
extinction coefficient derived from the Balmer decrements are similar and 
small for all the regions ($A_{V}$ ranges from 0 to 0.17). The oxygen 
abundances do not show significant variations from knot to knot. 

\item  The observed flux in the SF knots is the sum of, basically, three
components: the light coming from the young stars, the contribution of
emission lines and the emission from the low surface brightness
underlying population of stars. We presented a methodology to disentangle and
assess each of these components.

\begin{itemize}

\item Using our spectroscopic information we first computed the 
contribution of emission lines to the $B$, $V$ and $R$ filters, along the two
slit positions. We found that they can strongly affect the broad-band 
magnitudes and colors, as much as 0.7 mag in the $V$-band filter.
This contribution also varies significantly from one knot to the next. 
Thus, only by a two-dimensional mapping of the gas emission (which can be done by
means of Integral Field Spectroscopy) can we derive accurate
photometric values of the young star colors. 

\item We modeled the host galaxy with a \Sersic\ function, and subtracted it 
from the total emission to recover the colors of the young SF knots.

\end{itemize}

\item We compared the final observables with the predictions of
evolutionary synthesis models, and found that we can reproduce the colors
of the knots with an IB of star-formation and the Salpeter Initial Mass
Function with an upper mass limit of 100 M$_{\odot}$. The ages of the 
knots are about 5 Myr, with B being slightly older.
No significant age gradients have been found in the starburst region, which
discards the propagated star formation scenario.\newline

\item We found some evidence for the presence of dust in the central regions
of the galaxy. 

\item The kinematical analysis shows an overall rotation, with a peak-to-peak
amplitude of $\sim$ 100~km~s$^{-1}$. However, the curve is irregular in the
inner region, where there seems to be a counter-rotating component. This could
support the idea that the galaxy has undergone or is undergoing an interaction
or accretion process.

\end{itemize}

\acknowledgments

Based on observations with the WHT, operated by the Royal Greenwich
Observatory, and with the Nordic Optical Telescope, operated jointly by 
Denmark, Finland, Iceland, Norway, and Sweden, both  on the island of La Palma
in the Spanish Observatorio del Roque de los Muchachos of the Instituto de
Astrof{\'\i}sica de Canarias. We thank J.~N. Gonz{\'a}lez-P{\'e}rez for his
help in the initial stages of this project. We also thanks J.~M.
V{\'\i}lchez, J. Garc{\'\i}a-Rojas, A.~R. L{\'o}pez S{\'a}nchez and J.~H.
Huang for useful discussions. We are grateful to the unknown referee whose
detailed review and many suggestions greatly helped us to improve this paper. 
This research has made use of the NASA/IPAC Extragalactic Database (NED), which
is operated by the Jet Propulsion Laboratory, Caltech, under contract with the
National Aeronautics and Space Administration.  L.~M.~Cair{\'o}s acknowledges
support from the Alexander von Humboldt  foundation. This work has been
partially funded by the spanish  ``Ministerio de Ciencia y Tecnolog{\'\i}a''
(grants AYA2001-3939 and PB97-0158).

\end{document}